\input harvmac
\input epsf
\noblackbox
\newcount\figno
\figno=0
\def\fig#1#2#3{
\par\begingroup\parindent=0pt\leftskip=1cm\rightskip=1cm\parindent=0pt
\baselineskip=11pt
\global\advance\figno by 1
\midinsert
\epsfxsize=#3
\centerline{\epsfbox{#2}}
\vskip 12pt
\centerline{{\bf Figure \the\figno} #1}\par
\endinsert\endgroup\par}
\def\figlabel#1{\xdef#1{\the\figno}}
\def\pano{\par\noindent}

\font\cmss=cmss10
\font\cmsss=cmss10 at 7pt
\def\half{{1\over 2}}
\def\rlx{\relax\leavevmode}
\def\inbar{\vrule height1.5ex width.4pt depth0pt}
\def\IC{\relax\,\hbox{$\inbar\kern-.3em{\rm C}$}}
\def\IR{\relax{\rm I\kern-.18em R}}
\def\IN{\relax{\rm I\kern-.18em N}}
\def\IP{\relax{\rm I\kern-.18em P}}
\def\ZZ{\rlx\leavevmode\ifmmode\mathchoice{\hbox{\cmss Z\kern-.4em Z}}
 {\hbox{\cmss Z\kern-.4em Z}}{\lower.9pt\hbox{\cmsss Z\kern-.36em Z}}
 {\lower1.2pt\hbox{\cmsss Z\kern-.36em Z}}\else{\cmss Z\kern-.4em Z}\fi}
\def\narrowplus{\kern -.04truein + \kern -.03truein}
\def\narrowminus{- \kern -.04truein}
\def\narrowminussub{\kern -.02truein - \kern -.01truein}

\def\o#1{\overline{#1}}

\def\cK{{\cal{K}}}
\def\R{{\cal{R}}}
\def\F{{\cal{F}}}

\def\M{{\cal M}}
\def\C{{\cal C}}

\def\cA{{\cal{A}}}
\def\cM{{\cal{M}}}
\def\th#1#2{\vartheta\bigl[{\textstyle{  #1 \atop #2}} \bigr] }

\def\ie{{\it i.e.}}


\lref\rpss{D. Fioravanti, G. Pradisi, A. Sagnotti, {\it Sewing
 constraints and non-orientable open strings}, Phys. Lett. {\bf B321}
(1994) 349, {\tt hep-th/9311183}; G. Pradisi, A. Sagnotti,
Ya.S. Stanev, {\it Planar duality in $SU(2)$ WZW models},
Phys. Lett. {\bf B354} (1995) 279, {\tt hep-th/9503207}; 
{\it The open descendants of non-diagonal $SU(2)$ WZW models},
Phys. Lett. {\bf B356} (1995) 230, {\tt hep-th/9506014}; 
{\it Completeness conditions for boundary operators in 2D conformal
field theory}, Phys. Lett. {\bf B381} (1996) 97, 
{\tt hep-th/9603097}.} 

\lref\rbs{M. Bianchi, A. Sagnotti, {\it On the systematics of open
string theories}, Phys. Lett. {\bf B247} (1990) 517.}

\lref\rlerdarev{A. Lerda, R. Russo, {\it Stable non-BPS states
in string theory: a pedagogical review}, {\tt hep-th/9905006}.}

\lref\rfsuss{W. Fischler, L. Susskind, {\it Dilaton tadpoles,
string condensates and scale invariance}, Phys. Lett. {\bf B171}
(1986) 383; {\it Dilaton tadpoles, string condensates and scale
invariance. 2}, Phys. Lett. {\bf B173} (1986) 262.} 

\lref\rks{S. Kachru, E. Silverstein, {\it 4d conformal theories and
strings on orbifolds}, Phys. Rev. Lett. {\bf 80} (1998) 4855, 
{\tt hep-th/9802183}; S. Kachru, J. Kumar, E. Silverstein, 
{\it Vacuum energy cancellation in a non-supersymmetric string},
Phys. Rev. {\bf D59} (1999) 106004, {\tt hep-th/9807076}; J.A. Harvey,
{\it String duality and non-supersymmetric strings}, Phys. Rev. 
{\bf D59} (1999) 026002, {\tt hep-th/9807213}.}

\lref\rsagob{A. Sagnotti, {\it Some properties of open string
theories}, {\tt hep-th/9509080}; {\it Surprises in open string
perturbation theory}, {\tt hep-th/9702093}.}

\lref\rnilles{J. Lauer, J. Mas, H.P. Nilles, {\it Duality
and the Role of Nonperturbative Effects on the World-Sheet},
Phys. Lett. {\bf B226} (1989) 251;
E.J. Chun, J. Lauer, J. Mas, H.P. Nilles, {\it Duality and 
Landau-Ginzburg Models}, Phys. Lett. {\bf B233} (1989) 141.}

\lref\rbgkl{R. Blumenhagen, L. G\"orlich, B. K\"ors, D. L\"ust, 
{\it Asymmetric Orbifolds, Noncommutative Geometry and Type I
Vacua}, {\tt hep-th/0003024}.}

\lref\rbg{R. Blumenhagen, L. G\"orlich, {\it Orientifolds of 
Non-Supersymmetric Asymmetric Orbifolds}, Nucl. Phys. {\bf B551}
(1999) 601, {\tt hep-th/9812158}.}

\lref\rbgka{R. Blumenhagen, L. G\"orlich, B. K\"ors, 
{\it Supersymmetric Orientifolds in 6D with D-Branes at Angles},
Nucl. Phys. {\bf B569} (2000) 209, {\tt hep-th/9908130}; 
{\it A New Class of Supersymmetric Orientifolds with D-Branes at
Angles}, {\tt hep-th/0002146}.}

\lref\rbgkb{R. Blumenhagen, L. G\"orlich, B. K\"ors, {\it
Supersymmetric 4D Orientifolds of Type IIA with D6-branes at Angles},  
JHEP {\bf 0001} (2000) 040, {\tt hep-th/9912204}.}

\lref\rba{C. Angelantonj, R. Blumenhagen, {\it Discrete Deformations in 
Type I Vacua}, Phys. Lett. {\bf B473} (2000) 86, 
{\tt hep-th/9911190}.} 

\lref\rprad{G. Pradisi, {\it Type I Vacua from Diagonal
$Z_3$-Orbifolds}, {\tt hep-th/9912218}.}

\lref\raadds{C. Angelantonj, I. Antoniadis, G. D'Appollonio, E. Dudas,
A. Sagnotti, {\it Type I vacua with brane supersymmetry breaking},
{\tt hep-th/9911081}, to appear in Nucl. Phys. {\bf B}.}

\lref\rads{I. Antoniadis, E. Dudas, A. Sagnotti, {\it Brane
supersymmetry breaking}, Phys. Lett. {\bf B464} (1999) 38, 
{\tt hep-th/9908023}.}

\lref\raaf{C. Angelantonj, I. Antoniadis, K. F\"orger, 
{\it Non-supersymmetric type I strings with zero vacuum energy},
Nucl. Phys. {\bf B555} (1999) 116, {\tt hep-th/9904092}.}

\lref\razb{C. Angelantonj, {\it Non-tachyonic open descendants of the
0B string theory}, Phys. Lett. {\bf B444} (1998) 309, 
{\tt hep-th/9810214}.}

\lref\rbfl{R. Blumenhagen, A. Font, D. L\"ust, {\it Tachyon free
orientifolds of type 0B strings in various dimensions},
Nucl. Phys. {\bf B558} (1999) 159, {\tt hep-th/9904069}.} 

\lref\rbk{R. Blumenhagen, A. Kumar, {\it A note on orientifolds and
dualities of type 0B string theory}, Phys. Lett. {\bf B464} (1999) 46, 
{\tt hep-th/9906234}.}

\lref\rkf{K. F\"orger, {\it On non-tachyonic $Z_N \times Z_M$
orientifolds of type 0B string theory}, Phys. Lett. {\bf B469} (1999)
113, {\tt hep-th/9909010}.}

\lref\rbmp{M. Bianchi, J.F. Morales, G. Pradisi, {\it Discrete
torsion in non-geometric orbifolds and their open string descendants},
Nucl. Phys. {\bf B573} (2000) 314, {\tt hep-th/9910228}.}

\lref\rpolcai{J. Polchinski, Y. Cai, {\it Consistency of open
superstring theories}, Nucl. Phys. {\bf B296} (1988) 91.}

\lref\rbgone{O. Bergman, M.R. Gaberdiel, {\it A non-supersymmetric
open string theory and S-duality}, Nucl. Phys.~{\bf B499} (1997) 183, 
{\tt hep-th/9701137}.}

\lref\rbgtwo{O. Bergman, M.R. Gaberdiel, {\it Stable non-BPS
D-particles}, Phys. Lett.~{\bf B441} (1998) 133, 
{\tt hep-th/9806155}.}

\lref\rgabste{M.R. Gaberdiel, B. Stefa\'nski, jr.: {\it Dirichlet
Branes on Orbifolds}, Nucl. Phys. {\bf B578} (2000) 58, 
{\tt hep-th/9910109}.} 

\lref\rsena{A. Sen, {\it Stable non-BPS bound states of BPS
D-branes}, JHEP {\bf 9808} (1998) 010, {\tt hep-th/9805019}.}

\lref\rsenb{A. Sen, {\it Tachyon condensation on the brane
antibrane system}, JHEP {\bf 9808} (1998) 012, {\tt hep-th/9805170}.}

\lref\rsenrev{A. Sen, {\it Non-BPS States and Branes in String
Theory}, {\tt hep-th/9904207}.}

\lref\rgabrev{M.R. Gaberdiel, {\it Lectures on Non-BPS Dirichlet
branes}, {\tt hep-th/0005029}.}

\lref\rgabsen{M.R. Gaberdiel, A. Sen. {\it Non-supersymmetric D-Brane
Configurations with Bose-Fermi Degenerate Open String Spectrum}, 
JHEP {\bf 9911} (1999) 008, {\tt hep-th/9908060}.}

\lref\raldura{G. Aldazabal, A.M. Uranga, {\it Tachyon-free
Non-supersymmetric Type IIB Orientifolds via Brane-Antibrane Systems}, 
JHEP {\bf 9910} (1999) 024, {\tt hep-th/9908072}.}

\lref\raiq{G. Aldazabal, L.E. Ibanez, F. Quevedo, {\it Standard-like
Models with Broken Supersymmetry from Type I String Vacua}, JHEP 
{\bf 0001} (2000) 031, {\tt hep-th/9909172}; {\it A D-Brane
Alternative to the MSSM }, {\tt hep-ph/0001083}.}

\lref\raiqu{G. Aldazabal, L. E. Ibanez, F. Quevedo, A. M. Uranga, 
{\it D-Branes at Singularities : A Bottom-Up Approach to the String
Embedding of the Standard Model}, {\tt hep-th/0005067}.}

\lref\rsug{S. Sugimoto, {\it Anomaly cancellations in the type I
D9-anti-D9 system and the USp(32) string theory},
Prog. Theor. Phys. {\bf 102} (1999) 685, {\tt hep-th/9905159}.}

\lref\rss{J. Scherk, J.H. Schwarz, {\it Spontaneous breaking of
supersymmetry through dimensional reduction}, Phys. Lett. {\bf B82}
(1979) 60; {\it How to get masses from extra dimensions}, 
Nucl. Phys. {\bf B153} (1979) 61;
E. Cremmer, J. Scherk, J.H. Schwarz, {\it Spontaneously
broken ${\cal N}=8$ supergravity}, Phys. Lett. {\bf B84} (1979) 83.} 

\lref\rssst{R. Rohm, {\it Spontaneous supersymmetry breaking in
supersymmetric string theories}, Nucl. Phys. {\bf B237} (1984) 553; 
C. Kounnas, M. Porrati, {\it Spontaneous supersymmetry breaking in
string theory}, Nucl. Phys. {\bf B310} (1988) 355; 
S. Ferrara, C. Kounnas, M. Porrati, F. Zwirner, {\it Superstrings
with spontaneously broken supersymmetry and their effective theories},
Nucl. Phys. {\bf B318} (1989) 75; 
C. Kounnas, B. Rostand, {\it Coordinate dependent compactifications
and discrete symmetries}, Nucl. Phys. {\bf B341} (1990) 641;
I. Antoniadis, C. Kounnas, {\it Superstring phase transition at
high temperature}, Phys. Lett. {\bf B261} (1991) 369;
E. Kiritsis, C. Kounnas, {\it Perturbative and non-perturbative
partial supersymmetry breaking: ${\cal N} = 4 \to {\cal N} = 2 \to
{\cal N} =1$}, Nucl. Phys. {\bf B503} (1997) 117, 
{\tt hep-th/9703059}.} 

\lref\rssop{I. Antoniadis, E. Dudas, A. Sagnotti, 
{\it Supersymmetry breaking, open strings and M-theory},
Nucl. Phys. {\bf B544} (1999) 469, {\tt hep-th/9807011}; 
I. Antoniadis, G. D'Appollo\-nio, E. Dudas, A. Sagnotti, {\it Partial
breaking of supersymmetry, open strings and M-theory},
Nucl. Phys. {\bf B553} (1999) 133, {\tt hep-th/9812118}; 
{\it Open descendants of $\ZZ_2 \times \ZZ_2$ freely acting
orbifolds}, Nucl. Phys. {\bf B565} (2000) 123, {\tt hep-th/9907184}.} 

\lref\rmb{A. Sagnotti, {\it Anomaly cancellations and open-string
theories}, {\tt hep-th/9302099}; 
G. Zwart, {\it Four-dimensional ${\cal N}=1$ $\ZZ_N \times \ZZ_M$
orientifolds}, Nucl. Phys. {\bf B526} (1998) 378, 
{\tt hep-th/9708040}; 
Z. Kakushadze, G. Shiu, S.-H.H. Tye, {\it Type IIB orientifolds,
F-theory, type I strings on orbifolds and type I - heterotic duality},
Nucl. Phys. {\bf B533} (1998) 25, {\tt hep-th/9804092};
G. Aldazabal, A. Font, L. E. Ibanez, G. Violero, {\it $D=4$, ${\cal
N} =1$ type IIB orientifolds}, Nucl. Phys. {\bf B536} (1998) 29,
{\tt hep-th/9804026}.}

\lref\rau{G. Aldazabal, A.M. Uranga, {\it Tachyon-free
non-supersymmetric type IIB orientifolds via brane-antibrane systems}, 
JHEP {\bf 9910} (1999) 024, {\tt hep-th/9908072}.}

\lref\rmag{C. Bachas, {\it A way to break supersymmetry},
{\tt hep-th/9503030}; 
M. Bianchi, Ya.S. Stanev, {\it Open strings on the Neveu-Schwarz
penta-brane}, Nucl. Phys. {\bf B523} (1998) 193, {\tt hep-th/9711069}.}

\lref\rabq{C. Angelantonj, {\it Comments on open-string orbifolds with
a non-vanishing $B_{ab}$}, Nucl. Phys. {\bf B566} (2000) 126, 
{\tt hep-th/9908064}.}

\lref\ranomtad{G. Aldazabal, D. Badagnani, L.E. Ibanez, A.M. Uranga,
{\it Tadpole versus anomaly cancellation in $D=4,6$ compact IIB
orientifolds}, JHEP {\bf 9906} (1999) 031, {\tt hep-th/9904071};
M. Bianchi, J.F. Morales, {\it Tadpoles and anomalies}, JHEP 
{\bf 0003} (2000) 030, {\tt hep-th/0002149}.}

\Title{\vbox{
 \hbox{CPHT-S058.0600}
 \hbox{LPTENS 00/21}
 \hbox{HUB--EP--00/25}
 \hbox{DAMTP-2000-53}
}}
{\vbox
{\centerline{Asymmetric Orientifolds, Brane Supersymmetry}
\bigskip\centerline{Breaking and Non-BPS Branes}
}}
\vskip -.75cm
\centerline{Carlo Angelantonj\footnote{$^1$}{{\tt e-mail:
angelant@lpt.ens.fr}}$^{,a}$, 
Ralph Blumenhagen\footnote{$^2$}{{\tt e-mail:
blumenha@physik.hu-berlin.de}}$^{,b}$ and 
Matthias R. Gaberdiel\footnote{$^3$}{{\tt e-mail: 
M.R.Gaberdiel@damtp.cam.ac.uk}}$^{,c}$}
\medskip
\centerline{\it $^a$Laboratoire de Physique Th\'eorique de l'\'Ecole
Normale Superi\'eure\footnote{$^\dagger$}{Unit\'e mixte du CNRS et de
l'ENS, UMR 8549}}
\centerline{\it 24, rue Lhomond, 75231 Paris Cedex 05, France, and}
\centerline{\it Centre de Physique Th\'eorique, \'Ecole
Polytechnique\footnote{$^\ddagger$}{Unit\'e mixte du CNRS et de l'EP,
UMR 7644}, F-91128 Palaiseau} 
\centerline{\it $^b$Humboldt-Universit\"at zu Berlin, Institut f\"ur  
Physik,}
\centerline{\it Invalidenstrasse 110, 10115 Berlin, Germany}
\centerline{\it $^c$Department of Applied Mathematics and
Theoretical Physics}
\centerline{\it Centre for Mathematical Sciences,}
\centerline{\it Wilberforce Road, Cambridge CB3 0WA, U.K.}
\medskip
\centerline{\bf Abstract}
\noindent
A new class of six-dimensional asymmetric orientifolds is considered
where the orientifold operation is combined with T-duality. The models
are supersymmetric in the bulk, but the cancellation of the tadpoles
requires the introduction of brane configurations that break
supersymmetry. These can be described by  D7-brane anti-brane pairs,
non-BPS D8-branes or D9-brane anti-brane pairs. The transition between
these different configurations and their stability is analysed in
detail. 
\smallskip
\Date{06/2000}

\newsec{Introduction}

In the course of the last two years various attempts at constructing
non-super\-sym\-met\-ric open string models have been undertaken
\refs{\rbs,\rsagob,\rssop,\razb,\rbg,\rbfl,\raaf,\rsug,\rbk,\rads,
\raldura,\rabq,\rkf,\raiq,\raadds,\rbmp,\raiqu}.
These models always describe consistent string compactifications, but
the stability of the resulting theories is often difficult to 
establish. In particular, the theories often develop tachyonic
modes in certain regions of the moduli space. This indicates an
instability of the system to decay into another, quite possibly
supersymmetric, configuration.

Non-supersymmetric tachyon-free models in various space-time
dimensions have been constructed using a variety of different
approaches. In one approach one starts with the ten-dimensional
(non-supersymmetric) tachyonic type 0B string theory and performs a
suitable orientifold projection that removes the closed-string 
tachyon \rsagob. The various projections correspond to inequivalent
choices of the Klein bottle amplitude consistent with the constraints
of \rpss. Tadpole cancellation then requires the introduction
of D-branes, whose open string spectrum is also non-supersymmetric; in
the resulting theory supersymmetry is therefore broken in the bulk as
well as on the branes. It is also possible to consider
compactifications of these models \refs{\razb,\rbfl,\rbk,\rkf}, where
the orientifold projection is combined with an action of the space-time
group.  

An alternative way to break supersymmetry is via Scherk-Schwarz
compactifications \rss. In the simplest case of a circle
compactification, higher dimensional fields are allowed to be periodic
up to an R-symmetry transformation. Modular invariance suggests a
suitable extension of this mechanism to closed strings \rssst\ and
then, via orientifolds, to open strings \rssop\ as well. As a result
supersymmetry is broken both in the bulk and on the
branes\footnote{$^\dagger$}{Actually, in the M-theory construction, 
supersymmetry is still present at the massless level, although it is
broken for the massive modes.}, and different cosmological constants
are generated both in the bulk and on various branes. An interesting
variant of Scherk-Schwarz compactifications involving asymmetric
$\ZZ_2$ orbifold projections \rks\ leads to a non-supersymmetric
spectrum with Fermi-Bose degeneracy at all mass levels. Non-abelian
gauge symmetries can be generated via orientifolds \refs{\rbg,\raaf};
this leads to models where supersymmetry is preserved on the branes
(at lowest order), but is absent in the bulk. 

In a different approach, named in \rads\ {\it brane supersymmetry
breaking}, supersymmetry is broken only in the unoriented open-string
sector where different orientifold planes and D-branes are suitably
combined. In the simplest ten-dimensional case \rsug\ the orientifold
projection involves $O_-$ planes with positive tension and positive 
R-R charge instead of the more familiar $O_+$ with negative tension
and negative R-R charge. The projection on the closed string spectrum
is insensitive to the particular orientifold plane involved and
therefore yields the standard closed sector of the type I string, but
the cancellation of the massless R-R tadpoles now requires the
introduction of anti-branes with negative charge and positive
tension. Thus, the orientifold projection for the open-string bosons
is reverted and leads to a $USp(32)$ gauge group while the spinors are
still in the antisymmetric representation, consistently with anomaly
cancellation. As a result, a positive cosmological constant is
generated on the branes thus reflecting the impossibility to cancel
the NSNS tadpole. In this construction the choice of the types of
orientifold planes is optional and two different open-string spectra
(the supersymmetric $SO(32)$ and the non-supersymmetric $USp(32)$) can
both be consistently tied to a single closed-string spectrum. In lower
dimensional models, however, {\it brane supersymmetry breaking} is
often demanded by the consistency of the construction \rads\ and
represents a natural solution \raadds\ to old problems in
four-dimensional open-string model building \rmb. Further
non-supersymmetric deformations affecting only the open-string sector
involve a background magnetic field \rmag. 

Finally, in all (supersymmetric and non-supersymmetric) orientifold 
models one has the additional option to add pairs of branes and
antibranes consistently with tadpole cancellations
\refs{\rau,\raiq,\raiqu,\raadds}. As a result the bulk is not
affected, while supersymmetry (if present) is broken on the branes
where a cosmological constant is generated. Using similar
deformations, quasi-realistic theories with three generations in the
standard model, left-right symmetric extensions of the standard model
or Pati-Salam gauge groups have been constructed
\refs{\raiq,\raiqu}. Depending on the concrete model, the string scale
can be in the TeV range, or at an intermediate scale; the latter is
typical for models featuring gravity mediated supersymmetry breaking.

For models containing parallel branes and anti-branes of the same
dimension, the stability of the resulting configuration is
problematic. In particular, the system develops a tachyon (and thus
becomes unstable) if the branes and anti-branes come close together. 
In order to obtain a stable configuration it is therefore necessary to
remove the moduli that describe the relative distance between the
brane and the anti-brane. This can partially be achieved by
considering so--called fractional D-branes which are trapped at the
fixed points of some orbifold. However, the theory typically still
contains bulk moduli that describe the separation between the
different fixed points, and it is therefore necessary to remove those  
moduli as well. In all examples that have been studied so far, this 
could only be achieved dynamically, and the details were out of reach
of concrete computations.

In this paper we study a six dimensional orientifold, where we
combine the world-sheet parity reversal with T-duality, or rather
T-duality together with the symmetric reflection of two
directions.\footnote{$^\ddagger$}{These two descriptions are T-dual to
one another.} (Similar models were also considered in \rbmp.)
As we shall show, the Klein-bottle amplitude only leads to a twisted
R-R sector tadpole in our case. This is of significance since it
implies that the tadpoles can not be canceled by {\it any}
supersymmetric brane configuration: every BPS brane carries untwisted
RR charge, but since the entire D-brane configuration must have
vanishing untwisted R-R charge, it will {\it necessarily} also involve
anti-branes and therefore break supersymmetry. Thus one should expect
that the resulting non-supersymmetric theory (for a suitable brane 
configuration) is stable, and this is indeed what we shall find.  

In fact we shall find different brane configurations that cancel the
tadpoles (as well as the six-dimensional gravitational anomaly), and
we shall be able to understand how they can decay into one
another. The simplest solution consists of fractional D7-brane
anti-brane pairs where the D7-branes and the anti-D7-branes are
localized at different fixed points of the underlying
orbifold.\footnote{$^\star$}{The orientifold group is $\ZZ_4$, and it
contains a $\ZZ_2$ orbifold.} If the underlying torus is an orthogonal
torus at the $SU(2)^4$ point without any $B$-field, the ground state
of the string between the brane anti-brane pairs is either massless or
massive. Most of the bulk moduli that describe the shape of the torus
are removed by the orientifold projection (T-duality is only a
symmetry for a specific class of torii). However, certain shear
deformations remain, and if the torus is deformed in this way, an open
string tachyon develops in the string between some D7-brane anti-brane
pair. This tachyon indicates the instability of the brane anti-brane
system to decay into a non-BPS D8-brane with magnetic flux. The
resulting configuration of non-BPS D8-branes can be described in
detail and it also cancels the tadpoles (as well as the irreducible
gravitational and gauge anomalies); as far as we are aware, this is
the first time that non-BPS D-branes have naturally appeared in the
tadpole cancellation of an orientifold model.    

The configuration of non-BPS D8-branes still contains massless scalars
in the open string spectrum, and these can indeed become tachyonic if
another bulk modulus is turned on. The system then decays into a
configuration involving D9-branes and anti-D9-branes where both branes
and anti-branes carry magnetic flux of appropriate type. This can also
be constructed in detail, and indeed cancels the tadpoles as well as
the irreducible anomalies. The configuration is stable in a certain 
domain of the moduli space. However, if the torus is tilted
sufficiently, yet another tachyonic mode appears in the open string
spectrum, and the system decays into a configuration of diagonal
D7-brane anti-brane pairs; this final configuration appears to be
stable.

The paper is organized as follows. We begin in section~2 by describing
the model and the Klein bottle amplitude. Sections~3-6 deal with 
various brane configurations that cancel the R-R tadpoles. In
section~7 we discuss the stability of the different configurations,
and their deformations into one another. Finally, section~8 contains
some conclusions. We have included an appendix where the more
technical  material referring to the construction of boundary and
crosscap states is discussed.

\newsec{The definition of the model and the Klein bottle amplitude}
\pano

The model that we shall discuss in this paper is the asymmetric
orientifold of Type IIB string theory compactified on a 4-torus, whose
coordinates are labeled by $x_6,\ldots, x_9$. The orientifold group is
$\ZZ_4$, and it is generated by $\Omega \Theta_4$, where $\Omega$
denotes the standard Type IIB orientifold, and $\Theta_4$ describes
T-duality of the 4-torus. For the orthogonal self-dual $SU(2)^4$ torus
with vanishing internal B-field, T-duality is equivalent to the
asymmetric $\ZZ_2$ operation $I_4^L$. In the following we shall
discuss mostly this case, and we shall then use the notation $I_4^L$
for T-duality.   

The orientifold group contains a $\ZZ_2$ orbifold subgroup that is
generated by $I_4$; thus the theory is equivalently described as a
$\ZZ_2$ orientifold of Type IIB on K3. The $\Omega \Theta_4$ symmetry
fixes the metric of the $T^4$ completely, whereas the six independent
values of the internal $B$-field are free parameters. 

It is actually more convenient to consider the theory that is obtained
from the above after T-duality in the $x_7$ and $x_9$ directions,
say. If we denote by $\R$ the (symmetric) reflection in these two
coordinates, the theory in question is then described by 
\eqn\oriena{ {{\rm Type\, IIB}\ {\rm on}\ SU(2)^4\over 
            \{ (1+I_4)+\Omega \Theta_4 {\cal R} (1+I_4)  \}}\,. }
Under this T-duality, some of the $B$-field moduli of the original
model are mapped to geometric moduli of \oriena. In fact, looking
at the massless modes that survive the projection, one finds that
infinitesimally the following deformations are allowed
\eqn\metrbfield{ \delta G=\left(\matrix{ 0 & \delta g_{67} & 0 & 
\delta g_{69} \cr
      \delta g_{67} & 0 & \delta g_{78} & 0 \cr
      0 & \delta g_{78} & 0 & \delta g_{89} \cr
      \delta g_{69} & 0 & \delta g_{89} & 0 \cr}\right)\,,
\qquad 
      B=\left(\matrix{ 0 & 0 & \delta b_{68} & 0 \cr
              0 & 0 & 0 & \delta b_{79} \cr
             -\delta b_{68} & 0 & 0& 0 \cr
              0 & -\delta b_{79} & 0 & 0\cr}\right)\,. }
For simplicity we shall mainly consider the moduli that preserve the
decomposition of $T^4$ as $T^4=T^2\times T^2$, where the first $T^2$
has coordinates $x_6,x_7$, and the second has coordinates
$x_8,x_9$. (These moduli correspond to $g_{67}$ and $g_{89}$.) 
Let us determine to which global deformation of the torus these moduli
correspond to. The
action of the various symmetries on the complex structure $U$ and the
K\"ahler structure $T$ of $T^2$ is given as                               
\eqn\act{\eqalign{ &\Omega:(U,T)\mapsto (U, -\o{T}), \cr
                   &{\cal R}: (U,T)\mapsto (-\o{U}, -\o{T}),\cr
                   & \Theta_4: (U,T)\mapsto ( -1/U,-1/T)\,, \cr}}
so that the combined action is 
\eqn\comb{ \Omega{\cal R} \Theta_4:  (U,T)\mapsto (1/\o{U}, -1/T)\,.}
Thus $\Omega{\cal R} \Theta_4$ leaves a given $T^2$ invariant provided
that $T=i$ and $|U|^2=1$; this gives indeed rise to a one-dimensional
moduli space. 

Every two-torus can be described as the quotient of the complex plane,
$T^2=\IC / \Lambda$, where $\Lambda$ is a lattice. The tori that have
a complex structure $U$ satisfying $|U|=1$ are characterized by the
property that $\Lambda$ is generated by the basis vectors
\eqn\basis{ e_1={1\over R}\,,\quad\quad e_2={\kappa\over R} + i R \,,}
where $\kappa^2+R^4=1$. (More precisely $U$ is a phase, 
$U={\rm exp}(i\phi)$, and $\phi$ is the angle between $e_1$ and
$e_2$.) Furthermore, $T=i$ implies that the $B$-field vanishes. 
The torus is shown in figure 1.
\fig{\ The shape of the torus.}{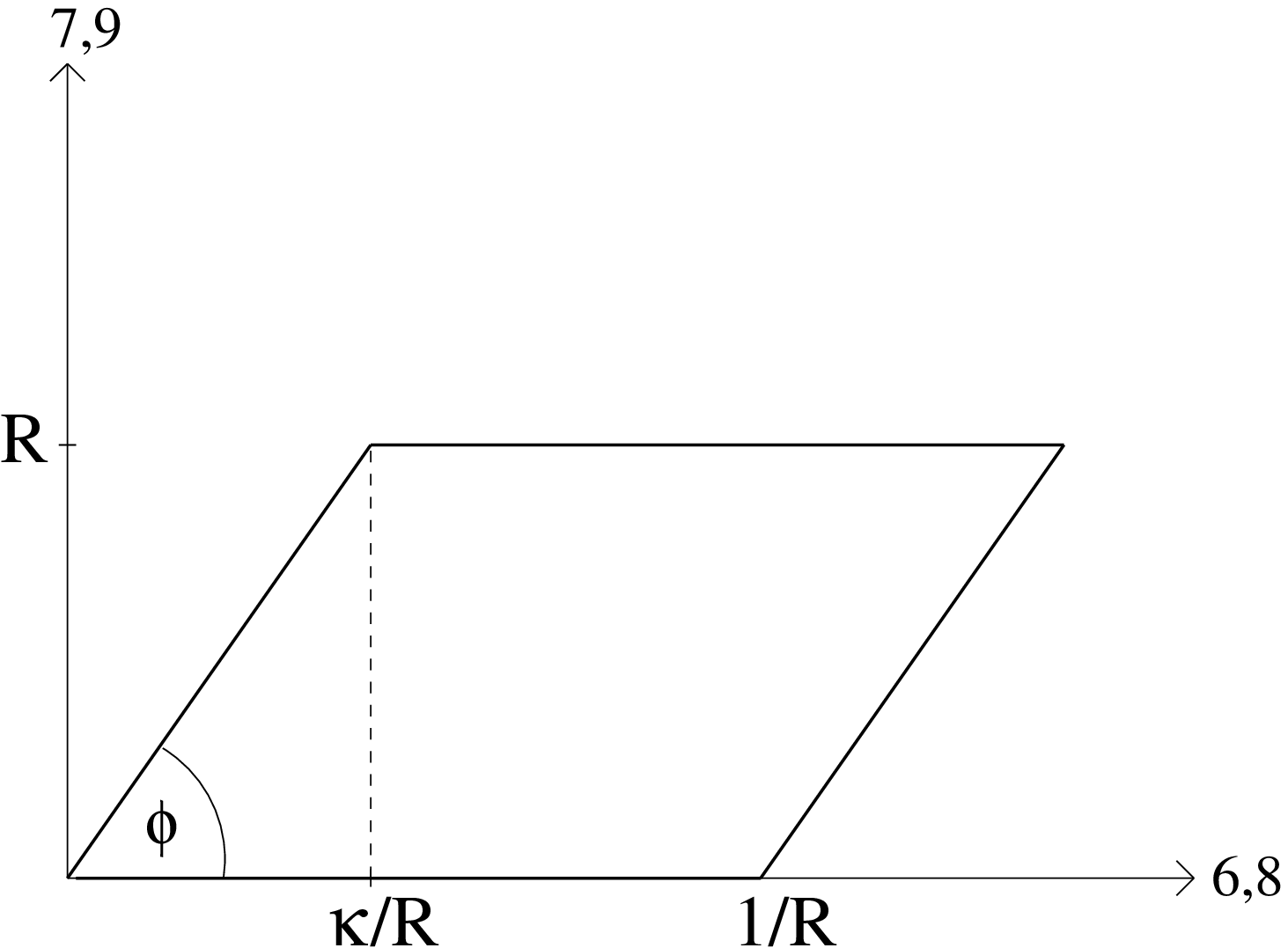}{8truecm}
We can write the left- and right-moving momenta as 
\eqn\zerom{\eqalign{ p_L&={1\over i\sqrt{U_2\, T_2}}
 \left[ U\, m_1-m_2- \o{T}\, (n_1+U\, n_2) \right] \cr
 p_R&={1\over i\sqrt{U_2\, T_2}}\left[ U\, m_1-m_2-
       {T}\, (n_1+U\, n_2) \right]\,, \cr }}
and we can thus directly determine the action of 
$\Omega{\cal R} \Theta_4$ on them; this leads to 
\eqn\momn{  \Omega{\cal R} \Theta_4:\cases{ p_L\to -{i {U}}\,\o{p}_R & 
\cr   p_R\to {i {U}}\,\o{p}_L\,. &\cr}}
Note, that the relation $(\Omega{\cal R} \Theta_4)^2=I_4$ holds in
general, even for $U\ne i$.

\subsec{The Klein bottle amplitude and the crosscap state}
\pano
First we study the Klein-bottle amplitude for the model \oriena\
defined on the $SU(2)^4$ torus. We want to determine the correct
crosscap states from the Klein bottle amplitude, which can be directly
computed in the loop channel; this is given by
\eqn\klein{ \cK =8\, \C \int_0^\infty {dt\over t^4}\, {\rm Tr}_{1,I_4}\,
\left( {\Omega{\cal R} I_4^L + \Omega{\cal R} I_4^R \over 4}\,
                P_{GSO}\, e^{-2\pi t (L_0 + \o{L}_0 )} \right) \,,}
where $\C=V_6/(8\pi^2\alpha')^3$ and the momentum integration over the
non-compact directions has already been performed. In the untwisted
sector we get 
\eqn\kleinu{ {\rm Tr}_{1}(\ldots) = { f_3^4\, f_4^4 - f_4^4\, f_3^4-
             f_2^4\, f_0^4 + f_0^4\, f_2^4\over f_1^4\, f_2^4 }}
with argument $q={\rm exp}(-2\pi t)$. The various $f$-functions are
defined by  
\eqn\tthet{\eqalign{ 
f_0(q)&= \sqrt{2}\,
q^{{1\over 12}} \prod_{n=0}^\infty (1 - q^{2n})=0 \,,
\cr
f_1(q)&=  q^{{1\over 12}} \prod_{n=1}^\infty (1 - q^{2n})\,, \cr
f_2(q)&= \sqrt{2}\, q^{{1\over 12}} \prod_{n=1}^\infty (1 + q^{2n})\,,
\cr
f_3(q)&= q^{-{1\over 24}} \prod_{n=1}^\infty (1 + q^{2n-1})\,, \cr 
f_4(q)&= q^{-{1\over 24}} \prod_{n=1}^\infty (1 - q^{2n-1})\,, }}
where we have used the notation of \rpolcai.
In the $I_4$ twisted sector the trace vanishes, as the action of
T-duality $I_4^L$ on the sixteen fixed points is given by the
traceless matrix \rnilles 
\eqn\fixed{ M=\bigotimes_{i=1}^4 {1\over \sqrt{2}} \left(
             \matrix{ 1 & 1 \cr 1& -1 \cr}\right)\,. }
After a modular transformation to the tree channel, writing
$t=1/(4l)$ and using the transformation rules of the $f_i$ functions, 
\eqn\ftrans{\eqalign{
f_0(e^{-\pi/l}) & = f_0(e^{-\pi l}) \,, \hskip20pt 
f_1(e^{-\pi/l}) = \sqrt{l} f_1(e^{-\pi l})\,, \cr
f_2(e^{-\pi/l}) & = f_4(e^{-\pi l})\,, \hskip20pt 
f_3(e^{-\pi/l}) = f_3(e^{-\pi l})\,,}}
we get
\eqn\kleintree{ \tilde \cK= \C\, \int_0^\infty dl\ 128\,  
        { f_3^4\, f_2^4 - f_4^4\, f_0^4-
             f_2^4\, f_3^4 + f_0^4\, f_4^4\over f_1^4\, f_4^4 }}         
with argument $q={\rm exp}(-2\pi l)$. It is immediate from this
expression that \kleintree\ only contains a twisted sector
tadpole. This is in contrast to ordinary $\ZZ_2$-orientifolds, where
only the untwisted sector propagates between the two crosscap
states. The fact that in our case only the twisted sector, $g=I_4$, is
allowed to flow between the two crosscap states is a consequence of
the relation $g=(\Omega{\cal R}I_4^L)^2=(\Omega{\cal R} I_4^R)^2$.   
The relevant crosscap states, $|\Omega{\cal R} I_4^{L,R}\rangle$ are
characterized by the equation 
\eqn\cross{ \left( X^\mu(\sigma,0) - {\cal R}\, I_4^{L,R}\, 
      X^\mu(\sigma+\pi,0)\right) |\Omega{\cal R} I_4^{L,R}\rangle=0 }
together with a similar condition for the world-sheet fermions; the
solution to these equations is constructed in the appendix. 

Under $I_4^L$ (or $I_4^R$) the two different crosscap states,
$|\Omega \R I_4^L\rangle$ and $|\Omega \R I_4^R\rangle$, are mapped
into one another. This follows from the identity 
\eqn\matj{ I_4^L (\Omega \R I_4^{L,R}) = (\Omega \R I_4^{R,L}) I_4^L
\,.} 
Thus the physical crosscap state is the sum of 
$|\Omega\R I_4^L\rangle$ and $|\Omega \R I_4^R\rangle$. Also, since
$I_4^L$ (or $I_4^R$) maps the sixteen fixed points non-trivially into
one another, the crosscap state must involve coherent states in 
different twisted sectors; the most symmetric solution involves then
all sixteen fixed points equally, and this is what we shall consider
in the following. Finally, the crosscap state is constrained by the
condition that the overlap with itself reproduces the tree level Klein
bottle amplitude \kleintree. This requires that it involves components
both from the twisted NS-NS and the twisted R-R sector, and that we
have 
\eqn\bedc{ \langle C_L|\, e^{-l\, H_{cl}}\,
    |C_R\rangle =0 \,,}
where $|C_L\rangle$ and $|C_R\rangle$ denote the total crosscap
states. A solution to all of these constraints is given by 
\eqn\twista{\eqalign{ |C_L\rangle=&(|L_1\rangle+|L_3\rangle)\otimes 
    (|L_1\rangle+|L_3\rangle)+(|L_1\rangle+|L_3\rangle)\otimes
    (|L_4\rangle-|L_2\rangle)\cr
    &+ (|L_4\rangle-|L_2\rangle)\otimes(|L_1\rangle+|L_3\rangle)+
     (|L_4\rangle-|L_2\rangle)\otimes (|L_4\rangle-|L_2\rangle) \cr
  |C_R\rangle=&(|R_1\rangle+|R_2\rangle)\otimes 
           (|R_1\rangle+|R_2\rangle)+
  (|R_1\rangle+|R_2\rangle)\otimes (|R_4\rangle-|R_3\rangle)\cr
     &+(|R_4\rangle-|R_3\rangle)\otimes (|R_1\rangle+|R_2\rangle)+
     (|R_4\rangle-|R_3\rangle)\otimes
    (|R_4\rangle-|R_3\rangle)\,. \cr}} 
Here 
\eqn\notaa{|L_i\rangle \otimes |L_j\rangle = 
     |\Omega\R I_4^L\rangle_{NSNS,T(ij)} 
          + |\Omega\R I_4^L\rangle_{RR,T(ij)}\,,}
where $|\Omega\R I_4^L\rangle$ is defined in the appendix, and the
twisted sector denoted by $T(ij)$ is localized at the fixed point
$T_i$ of the $T^2$ with coordinates $x_6,x_7$, and at the fixed point 
$T_j$ of the $T^2$ with coordinates $x_8,x_9$; on each $T^2$ we denote
the different fixed points by 
\eqn\fourfp{ T_1=\biggl(0,0\biggr),\ T_2=\left(\half,0\right),\ 
             T_3=\left(0,\half\right),\ 
             T_4=\left(\half,\half\right) \,.} 
The notation for $|R_i\rangle \otimes |R_j\rangle$ is analogous. 

Both $|C_L\rangle$ and $|C_R\rangle$ can be thought to consist of four
parallel O7-planes that `stretch' between four fixed points each (and 
fill the uncompactified space). For example 
$(|L_1\rangle+|L_3\rangle)\otimes (|L_1\rangle+|L_3\rangle)$ defines
an O7-plane that is localized at $x_6=x_8=0$ and `stretches' between
the four fixed points with $x_7=0,1/2$ and $x_9=0,1/2$; similar
statements also hold for the other terms in \twista. 

The ${\cal N}=(0,1)$ supersymmetric massless spectrum consists of the 
supergravity multiplet, 11 tensor multiplets and 10 hypermultiplets.
There exist various configuration of D-branes that cancel the twisted
sector tadpoles from the Klein bottle and therefore cancel the
anomalies from the closed string sector. These configurations will be
discussed in turn.

\newsec{The D7-brane antibrane configuration}
\pano
The simplest configuration of branes that cancels the tadpoles
consists of D7-branes and anti-branes that are arranged in the same
way as the O7-planes. The branes in question are so--called
`fractional' branes whose boundary states have components in the
untwisted as well as the twisted sectors. In particular, this implies
that the branes are stuck at the fixed planes. Since the Klein bottle
amplitude does not have any untwisted R-R tadpoles, we have to
introduce D7-branes and D7-antibranes ($\o{{\rm D}7}$) in pairs. 

The boundary states of the D7-branes are schematically described by
(see for example \rgabrev\ for an introduction into these matters)
\eqn\bounda{ |D7\rangle= 
\left( |U,NS\rangle+|U,R\rangle\right) 
+ \left( |T,NS\rangle+ |T,R\rangle\right) \,,} 
where $U$ denotes the untwisted sector and $T$ the twisted sector.
The normalization factors for the untwisted and twisted sector
parts have to be determined by loop channel-tree channel equivalence.
Under the action of $\Omega{\cal R}I_4^L$, a $D7$-brane is mapped to
a $\widetilde{D7}$ brane that is orthogonal to the former. In
particular this implies that the tree exchange between a $D7$-brane  
and a $\widetilde{D7}$ brane vanishes. The boundary states for the
antibranes, $\o{D7}$, are of the form
\eqn\boundb{ |\o{D7}\rangle=
\left( |U,NS\rangle-|U,R\rangle\right) 
- \left( |T,NS\rangle-|T,R\rangle\right) \,,} 
and thus the open string that stretches between a $D7$ and a $\o{D7}$
brane has the opposite GSO projection. (It also has the opposite $I_4$
projection.) In order to achieve a local cancellation of the tadpoles
the charge assignment for the various twisted sector ground states
must be chosen as for the O7-planes in \twista. 

We also have to guarantee that the total configuration is invariant
under $I_4^{L,R}$, \ie\ under T-duality. Under this operation, spatial
distances are exchanged with Wilson lines which in turn are related to
the relative signs of the twisted (R-R) charges at different fixed
points. A consistent choice for the $D7$-branes and $\o{D7}$-branes is
described in table~1 (see also figures~2 and 3). (The last column in
table~1 describes the action on the Chan-Paton factors for the term in
the open string with the insertion of $I_4$; this can be determined
from the given boundary states by world-sheet duality.) 
\vskip 0.2cm
\vbox{
\centerline{\vbox{
\hbox{\vbox{\offinterlineskip
\def\tablespace{height2pt&\omit&&\omit&&\omit&&\omit&&
 \omit&\cr}
\def\tablerule{\tablespace\noalign{\hrule}\tablespace}

\hrule\halign{&\vrule#&\strut\hskip0.2cm\hfil#\hfill\hskip0.2cm\cr
\tablespace
& brane && location  && Wilson line && twisted sector && 
                $(\gamma_{I_4})$ &\cr
\tablerule
& $D7_1$ && $x_6=0$, $x_8=0$ && $\theta_7=0$, $\theta_9=0$ &&
$(T_1+T_3)(T_1+T_3)$ && $I$ &\cr
\tablespace
 & $\o{D7}_2$ && $x_6=0$, $x_8={1\over 2}$ && $\theta_7=0$, 
$\theta_9={1\over 2}$  && $(T_1+T_3)(T_4-T_2)$ && $I$ &\cr
\tablespace
 & $\o{D7}_3$ && $x_6={1\over 2}$, $x_8=0$ && $\theta_7={1\over 2}$, 
$\theta_9=0$  && $(T_4-T_2)(T_1+T_3)$ && $I$ &\cr
\tablespace
 & ${D7}_4$ && $x_6={1\over 2}$, $x_8={1\over 2}$ && 
$\theta_7={1\over 2}$, $\theta_9={1\over 2}$  && 
$(T_4-T_2)(T_4-T_2)$ && $I$ &\cr
\tablerule
& $\widetilde{D7}_1$ && $x_7=0$, $x_9=0$ && $\theta_6=0$, 
$\theta_8=0$ && $(T_1+T_2)(T_1+T_2)$ && $I$ &\cr
\tablespace
 & $\widetilde{\o{D7}}_2$ && $x_7=0$, $x_9={1\over 2}$ &&
$\theta_6=0$, $\theta_8={1\over 2}$ && $(T_1+T_2)(T_4-T_3)$ && $I$
&\cr 
\tablespace
 & $\widetilde{\o{D7}}_3$ && $x_7={1\over 2}$, $x_9=0$ && 
$\theta_6={1\over 2}$, $\theta_8=0$  && $(T_4-T_3)(T_1+T_2)$ 
&& $I$ &\cr
\tablespace
 & $\widetilde{{D7}}_4$ && $x_7={1\over 2}$, $x_9={1\over 2}$ && 
$\theta_6={1\over 2}$, $\theta_8={1\over 2}$  && 
$(T_4-T_3)(T_4-T_3)$ && 
$I$ &\cr
\tablespace}\hrule}}}}
\centerline{
\hbox{{\bf Table 1:}{\it ~~ D7-branes for model I.}}}}
\vskip 0.5cm
\noindent
\fig{\ The location of the $D7$-branes.}{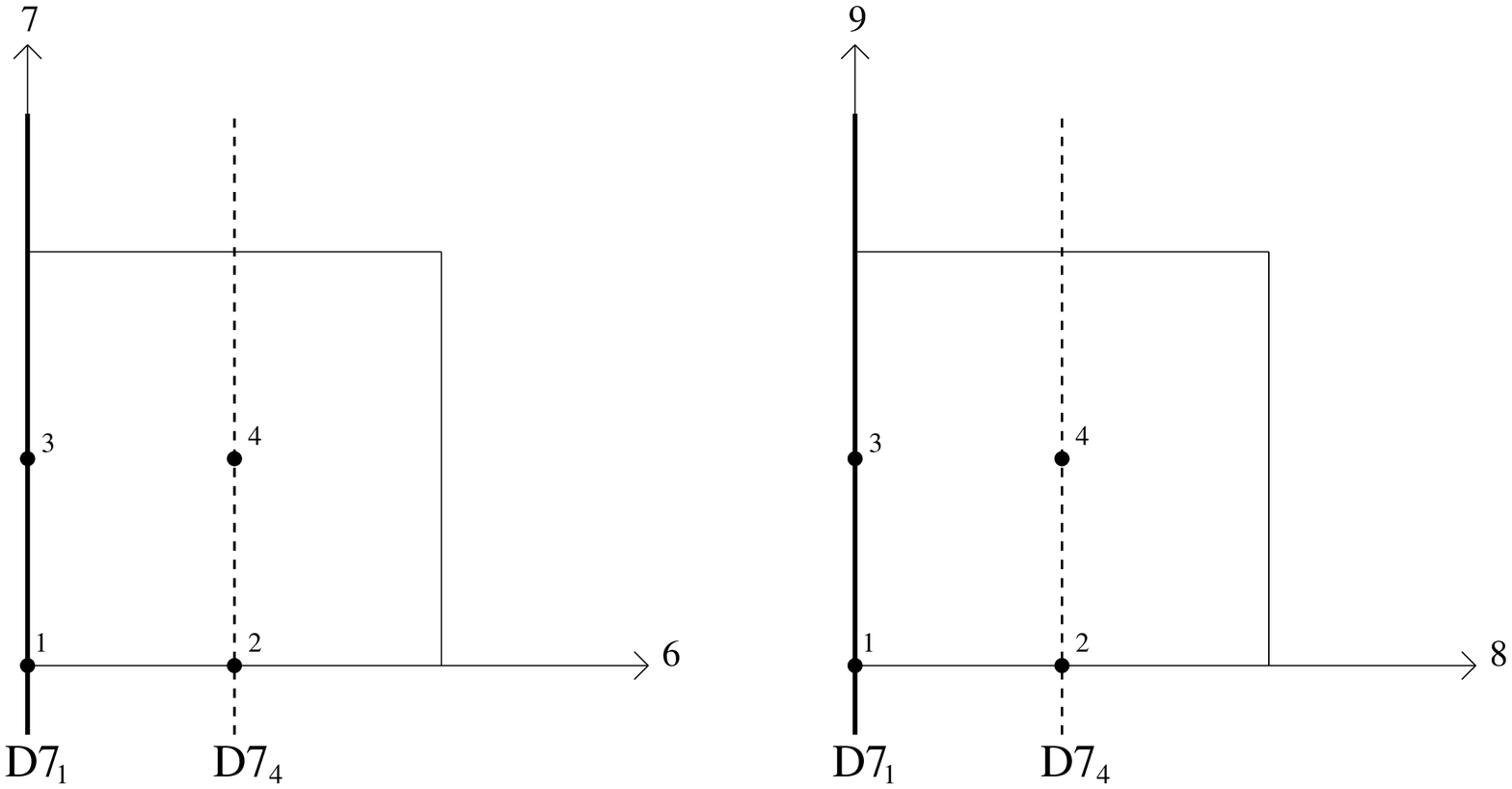}{12.5truecm}
\fig{\ The location of the $\o{D7}$-branes.}{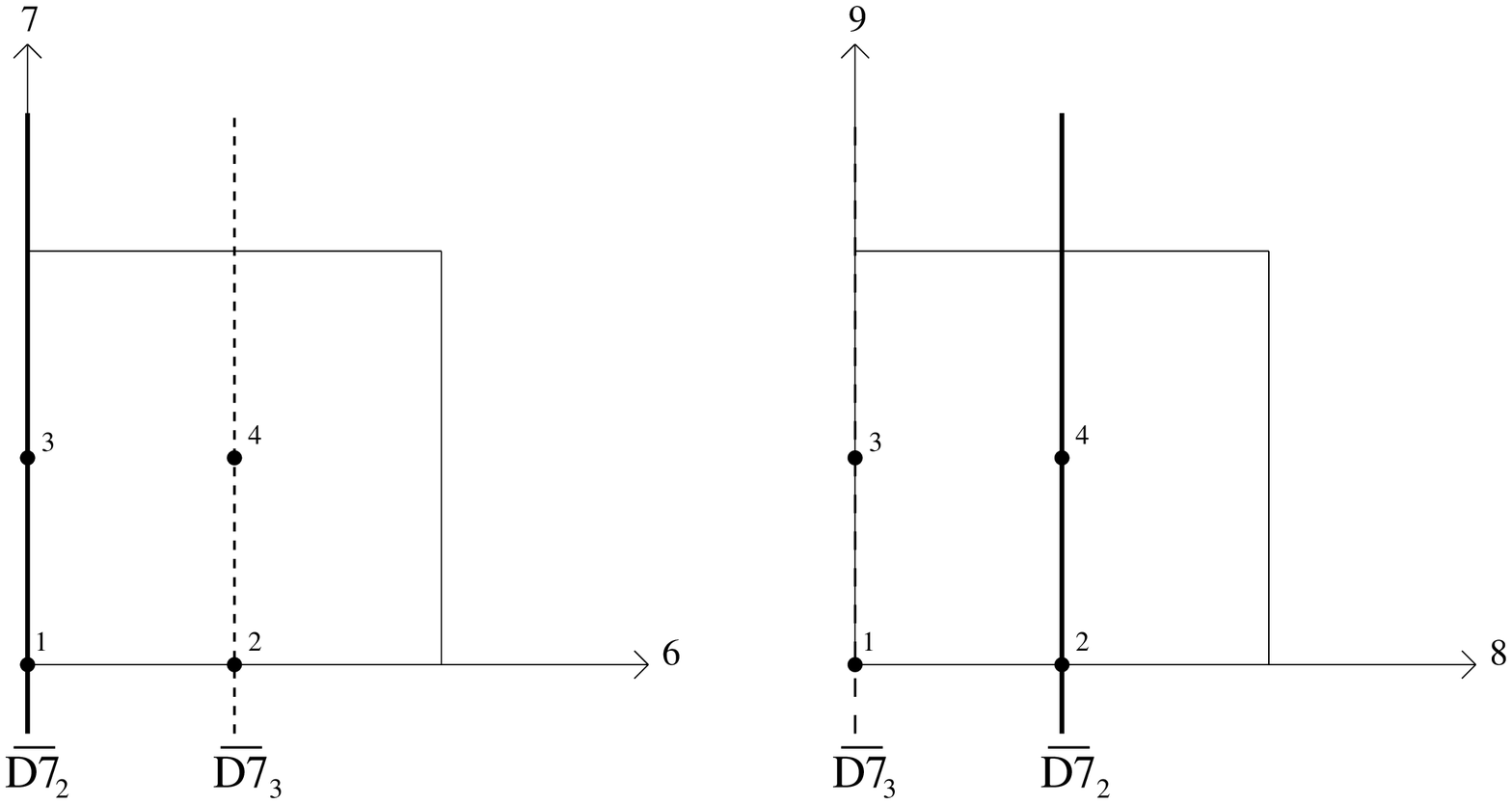}{12.5truecm}
\noindent
Given the branes described in table~1, it is straightforward to
compute the annulus amplitude in the tree channel, and the 
result is 
\eqn\anng{\eqalign{  \tilde{\cA}=\C \int_0^\infty dl\ 
{\rm N}^2 \Biggl\{
2&\left( {f_3^8-f_4^8 \over f_1^8 }\right)_{\rm NSNS,U} 
 \left( \Theta_{0,1}^2+\Theta_{1,1}^2\right)^2  
-8 \left( {f_2^8-f_0^8  \over f_1^8 }\right)_{\rm RR,U} 
   \Theta_{0,1}^2\, \Theta_{1,1}^2  \cr
+8&\left( {f_3^4 f_2^4 - f_4^4 f_0^4-f_2^4 f_3^4+f_0 f_4^4\over
             f_1^4 f_4^4} \right)_{\rm NSNS-RR,T} \cr
+8&\left( { f_3^4 f_4^4 -f_4^4 f_3^4 \over 
             f_1^4 f_2^4}\right)_{\rm NSNS,U}  
-8\left( { f_4^4 f_2^4-f_0^4 f_3^4 
\over f_1^4 f_3^4 }\right)_{\rm NSNS,T}
  \Biggr\}\,,\cr }}
where the argument is $q={\rm exp}(-2\pi l)$, and we have used the
standard definition for  
\eqn\tata{ \Theta_{j,k}(q^2)=\sum_{m\in\ZZ} 
q^{2k\left(m+{j\over 2k} \right)^2}\,. }
In the above, we have assumed that we have the same number, N, of each
of the D-branes. We should note that the massless untwisted
R-R-tadpole vanishes and that the fourth term in \anng\ does not give
rise to any massless tadpole, either. Upon world-sheet duality,
writing $t=1/(2l)$, the annulus amplitude becomes 
\eqn\annf{\eqalign{  \cA=\C\int_0^\infty &{dt\over t^4}\ {\rm N}^2\ 
\Biggl\{
\left( {f_3^8-f_4^8-f_2^8+f_0^8 \over f_1^8}\right) 
  \left( \Theta_{0,1}^4+\Theta_{1,1}^4 \right) \cr
&+ \left( {f_3^8+f_4^8-f_2^8-f_0^8 \over f_1^8} \right)\, 
 2\, \left( \Theta_{0,1}^2 \Theta_{1,1}^2 \right) 
+ 4\left( {f_3^4 f_4^4 - f_4^4 f_3^4-f_2^4 f_0^4+f_0 f_2^4\over
             f_1^4 f_2^4}\right) \cr
&
+ 4\left( { f_3^4 f_2^4 -f_2^4 f_3^4 \over f_1^4 f_4^4}\right)  
- 4\left( { f_2^4 f_4^4-f_0^4 f_3^4 \over f_1^4 f_3^4 }\right)
\Biggr\}\,, \cr 
}}
where the argument is $q={\rm exp}(-\pi t)$. Here we have used that
the $\Theta_{j,1}$ functions transform as 
\eqn\thtrans{\eqalign{ \Theta_{0,1}(e^{-2\pi/\tau}) & = 
\sqrt{\tau\over 2}\, \Bigl(\Theta_{0,1}(e^{-2\pi\tau}) 
                      + \Theta_{1,1}(e^{-2\pi\tau}) \Bigr)\,, \cr
\Theta_{1,1}(e^{-2\pi/\tau}) & = 
\sqrt{\tau\over 2}\, \Bigl(\Theta_{0,1}(e^{-2\pi\tau}) 
                      - \Theta_{1,1}(e^{-2\pi\tau}) \Bigr) \,.}}
The first term in \annf\ arises from open strings stretched between
parallel $D7$-branes or parallel $\o{D7}$-branes, without the
insertion of $I_4$ in the trace. The second term comes from open
strings stretched between parallel $D7$ and $\o{D7}$ branes without
$I_4$ insertion. The third term is due to parallel D-branes with $I_4$
insertion in the trace, the fourth term is from orthogonal D-branes
without $I_4$ insertion and finally the fifth term comes from
orthogonal branes with $I_4$ insertion.

The M\"obius strip amplitude is determined, in the tree channel, by
the overlap between the total crosscap state and the different
D7-brane states. The different contributions include for example
\eqn\moebi{\eqalign{ &\int_0^\infty dl\ 
\langle \Omega{\cal R} I_4^{L}| 
e^{-l H_{cl}}| D7_{1,4}\rangle=-\C\int_0^\infty dl\ 
4\, {\rm N}\ 
\left( { f_3^4 f_2^4 - f_4^4 f_0^4 - f_2^4 f_3^4+f_0^4 f_4^4 
   \over f_1^4 f_4^4}\right) \cr
     &\int_0^\infty dl\ \langle \Omega{\cal R} I_4^{R}| 
         e^{-l H_{cl}}| D7_{1,4}\rangle=0 \,, } }
where the argument is $q=i{\rm exp}(-2\pi l)$. Adding all
these terms together, we find that the contribution of the 
(twisted) R-R sector is given by
\eqn\moebfinal{ \widetilde{\cM}=\C\int_0^\infty dl\ 32\, {\rm N}\ 
      \left(    { f_2^4 f_3^4 -f_0^4 f_4^4\over f_1^4 f_4^4 } +
     { f_2^4 f_4^4 -f_0^4 f_3^4\over f_1^4 f_3^4 }\right)_{\rm RR,T}
     \,.}    
Note that \moebfinal\ is indeed real, as under complex conjugation one
has
\eqn\complc{\Bigl(f_3(iq)\Bigr)^*=e^{-{\pi i\over 24}} f_4(iq),\quad
            \Bigl(f_4(iq)\Bigr)^*=e^{-{\pi i\over 24}} f_3(iq)\,.}
In order to determine the massless spectrum of the open strings 
we have to determine the M\"obius strip amplitude in loop channel for
the different combinations separately, and we find that
\eqn\moebiloop{\eqalign{  \int_0^\infty {dt\over t^4} \,&
  {\rm Tr}_{1\tilde 1, 4\tilde 4}
\left(  {\Omega{\cal R} I_4^L + \Omega{\cal R} I_4^R \over 4}\,
                P_{GSO}\, e^{-2\pi t L_0  } \right) \cr
     &= {{\rm N}\over 2} \int_0^\infty {dt\over t^4}  \ 
\left( {f_3^4 f_2^4-f_4^4 f_0^4 -f_2^4 f_3^4 +f_0^4 f_4^4 
\over f_1^4 f_4^4} +        
{ f_4^4 f_2^4-f_3^4 f_0^4  -f_2^4 f_4^4 +f_0^4 f_3^4 
\over f_1^4 f_3^4}
\right) ,\cr}}
with argument $q=i{\rm exp}(- \pi t)$, so that for $D7$-branes both
the NS and the R sector are symmetrized. For the two pairs of
$\o{D7}$-branes we obtain 
\eqn\moebiloopb{\eqalign{ \int_0^\infty {dt\over t^4} \, &
  {\rm Tr}_{2\tilde 2, 3\tilde 3}
\left(  {\Omega{\cal R} I_4^L + \Omega{\cal R} I_4^R \over 4}\,
                P_{GSO}\, e^{-2\pi t L_0  } \right)  \cr
     & = {{\rm N}\over 2} \int_0^\infty {dt\over t^4}  \ 
\left( {f_4^4 f_0^4 -f_3^4 f_2^4 -f_2^4 f_3^4 
+f_0^4 f_4^4 \over f_1^4 f_4^4}
+        
{ f_3^4 f_0^4 -f_4^4 f_2^4 -f_2^4 f_4^4 +f_0^4 f_3^4 \over f_1^4 f_3^4}
\right) ,\cr}}
so that the NS sector is now antisymmetrized whereas the R-sector is
still symmetrized. The change of sign between $\cM$ and
$\widetilde{\cM}$ is due to 
\eqn\sign{   f_3 \to  f_4, \quad\quad
             f_4\to e^{-i\pi/4} f_3 }
under the $P=TST^2S$ transformation (that relates the tree and the
loop channel M\"obius strip amplitude).

All three tree-level diagrams do not have a massless untwisted R-R
tadpole, and therefore only the twisted (massless) R-R tadpole needs
to be canceled. It follows from \kleintree, \anng\ and \moebfinal\
that this requires 
\eqn\tada{ 8\, {\rm N}^2 -64\, {\rm N} +128 = 8\, ({\rm N}-4)^2=0 \,,}
so that we need four D-branes of each kind. On the other hand, we
can not cancel one untwisted and one twisted NS-NS-tadpole; this will
therefore lead to a shift in the background via the  Fischler-Susskind
mechanism \rfsuss.  

The massless spectrum in the closed string sector consists of the
${\cal N}=(0,1)$ supergravity multiplet in addition to 11 tensor
multiplets and 10 hypermultiplets, and the massless spectrum arising
from the different open strings is listed in table~2. 
\vskip 0.8cm
\centerline{\vbox{
\hbox{\vbox{\offinterlineskip
\def\tablespace{height2pt&\omit&&
 \omit&\cr}
\def\tablerule{\tablespace\noalign{\hrule}\tablespace}

\hrule\halign{&\vrule#&\strut\hskip0.2cm\hfil#\hfill\hskip0.2cm\cr
\tablespace
& spin && states &\cr
\tablerule
& $(2,2)$ &&  U(4)$\times$U(4)$\times$U(4)$\times$U(4) & \cr
\tablespace
& $(2,1)$ &&  $2\times\left[ ({\rm adj},1,1,1)+(1,{\rm adj},1,1)+
(1,1,{\rm adj},1)+ (1,1,1,{\rm adj}) \right]$ &\cr
\tablerule
& $(1,1)$ && $2\times [ (4,\o{4},1,1)+(\o{4},4,1,1)+(4,1,\o{4},1)+
(\o{4},1,4,1)] +$ &\cr
\tablespace
&   && $2\times [(1,4,1,\o{4})+(1,\o{4},1,4)+(1,1,4,\o{4})+
    (1,1,\o{4},{4}) ] $ &\cr
\tablerule
& $(1,2)$ && $\left[ (10+\o{10},1,1,1)+(1,10+\o{10},1,1)+
(1,1,10+\o{10},1)+(1,1,1,10+\o{10}) \right]$ &\cr
\tablerule
& $(1,1)$ && $2\times\left[ (10+\o{10},1,1,1)+(1,6+\o{6},1,1)+
(1,1,6+\o{6},1)+(1,1,1,10+\o{10}) \right]$ &\cr
\tablerule
& $(2,1)$ && $[ (4,4,1,1)+(\o{4},\o{4},1,1)+(4,1,4,1)+
(\o{4},1,\o{4},1)]+$ &\cr
\tablespace
&  && $[ (1,4,1,4)+(1,\o{4},1,\o{4})+(1,1,4,4)
+(1,1,\o{4},\o{4}) ] $ &\cr
\tablerule
& $(1,2)$ && $\left[ (4,1,1,4)+(\o{4},1,1,\o{4})+(1,4,4,1)+
(1,\o{4},\o{4},1) \right] $ &\cr
\tablerule
& $(1,1)$ && $2\times \left[ (4,1,1,4)+(\o{4},1,1,\o{4})+(1,4,4,1)+
(1,\o{4},\o{4},1) \right] $ &\cr
\tablespace}\hrule}}}}
\centerline{
\hbox{{\bf Table 2:}{\it ~~ Massless open string spectrum for model I
(orthogonal $D7$-branes).}}}
\vskip 0.5cm 
\noindent 
It is worth mentioning that the sector of open strings starting and
ending on the same D-brane does not contain any scalar moduli. This
implies that the D7-branes are not allowed to move off the fixed
points. This is basically a consequence of the fact that all four
fractional branes carry the {\it same} twisted R-R charge; pairs of 
fractional branes can move off a fixed point, but only if they carry
the opposite twisted charge.  

The distance between the different branes and anti-branes is such that
the ground state `tachyon' is either massless or massive; in
particular, the open string spectrum therefore does not contain any
actual tachyons. Since the branes and antibranes are fixed to lie on
the fixed points, their distance is determined in terms of the radii
of the underlying torus. However, these radii are not moduli any more 
since the theory is only well-defined for a self-dual torus, and the
configuration is (at least at this level) stable.

The non-supersymmetric spectrum of table~2 also contains $N_+=256$
massless fermions of $(2,1)$ chirality, and $N_-=144$ massless
fermions of $(1,2)$ chirality, giving rise to $\Delta N=N_+-N_-=112$;
this is precisely what is needed to cancel the non-factorizable
gravitational anomaly. Moreover, the configuration of table 2 is also
free of irreducible gauge anomalies, consistently with the R-R tadpole
cancellation \refs{\ranomtad}.  
\vskip 0.5cm

The arrangement of $D7$ and $\o{D7}$ branes listed in table~1 is not
the only possible configuration of parallel $D7$-branes and
anti-branes that cancels the R-R-tadpole. In fact, we can exchange the
roles of some of the branes and anti-branes, and consider instead the
brane configuration described in table~3. (This configuration differs
from that described in table~1 by the property that $D7_2$ is now a
brane, whereas $D7_4$ is now an anti-brane.)
\vskip 0.2cm
\centerline{\vbox{
\hbox{\vbox{\offinterlineskip
\def\tablespace{height2pt&\omit&&\omit&&\omit&&\omit&&
 \omit&\cr}
\def\tablerule{\tablespace\noalign{\hrule}\tablespace}

\hrule\halign{&\vrule#&\strut\hskip0.2cm\hfil#\hfill\hskip0.2cm\cr
\tablespace
& brane && location  && Wilson line && twisted sector && 
                $(\gamma_{I_4})$ &\cr
\tablerule
& $D7_1$ && $x_6=0$, $x_8=0$ && $\theta_7=0$, $\theta_9=0$ &&
$(T_1+T_3)(T_1+T_3)$ && $I$ &\cr
\tablespace
 & $D7_2$ && $x_6=0$, $x_8={1\over 2}$ && $\theta_7=0$, 
$\theta_9={1\over 2}$ 
 && $(T_1+T_3)(T_4-T_2)$ && $-I$ &\cr
\tablespace
 & $\o{D7}_3$ && $x_6={1\over 2}$, $x_8=0$ && $\theta_7={1\over 2}$, 
$\theta_9=0$  && $(T_4-T_2)(T_1+T_3)$ && $I$ &\cr
\tablespace
 & $\o{D7}_4$ && $x_6={1\over 2}$, $x_8={1\over 2}$ && 
$\theta_7={1\over 2}$, 
$\theta_9={1\over 2}$  && $(T_4-T_2)(T_4-T_2)$ && $-I$ &\cr
\tablerule
& $\widetilde{D7}_1$ && $x_7=0$, $x_9=0$ && $\theta_6=0$, 
$\theta_8=0$ &&
$(T_1+T_2)(T_1+T_2)$ && $I$ &\cr
\tablespace
 & $\widetilde{D7}_2$ && $x_7=0$, $x_9={1\over 2}$ && $\theta_6=0$, 
$\theta_8={1\over 2}$ && $(T_1+T_2)(T_4-T_3)$ && $-I$ &\cr
\tablespace
 & $\widetilde{\o{D7}}_3$ && $x_7={1\over 2}$, $x_9=0$ && 
$\theta_6={1\over 2}$, $\theta_8=0$  && $(T_4-T_3)(T_1+T_2)$ 
&& $I$ &\cr
\tablespace
 & $\widetilde{\o{D7}}_4$ && $x_7={1\over 2}$, $x_9={1\over 2}$ && 
$\theta_6={1\over 2}$, $\theta_8={1\over 2}$  && 
$(T_4-T_3)(T_4-T_3)$ && 
$-I$ &\cr
\tablespace}\hrule}}}}
\centerline{
\hbox{{\bf Table 3:}{\it ~~ D7-branes for model II.}}}
\vskip 0.5cm
\noindent
This modification only changes the annulus amplitude which, in the
tree channel, now becomes
\eqn\anngg{\eqalign{  \widetilde{\cA}=\C\int_0^\infty dl\ {\rm N}^2 
\Biggl\{
2&\left( {f_3^8-f_4^8\over f_1^8}\right)_{\rm NSNS,U} 
 \left( \Theta_{0,1}^2+\Theta_{1,1}^2\right)^2  \cr
-4& \left( {f_2^8-f_0^8 \over f_1^8 }\right)_{\rm RR,U}  
\left(\Theta_{0,1}^2+\Theta_{1,1}^2\right) 
   \Theta_{0,1}\, \Theta_{1,1}  \cr
+8&\left( {f_3^4 f_2^4 - f_4^4 f_0^4-f_2^4 f_3^4+f_0 f_4^4\over
             f_1^4 f_4^4} \right)_{\rm NSNS-RR,T} \cr
+8&\left( { f_3^4 f_4^4 -f_4^4 f_3^4 \over 
             f_1^4 f_2^4}\right)_{\rm NSNS,U}  
\Biggr\}\,,\cr }}
In particular, the contribution of the massless R-R sector states in 
\anngg\ is unmodified compared to \anng, and therefore ${\rm N}=4$
still cancels the R-R tadpoles. The massless spectrum of this model is
described in table~4. 
\vskip 0.8cm
\centerline{\vbox{
\hbox{\vbox{\offinterlineskip
\def\tablespace{height2pt&\omit&&\omit&&
 \omit&\cr}
\def\tablerule{\tablespace\noalign{\hrule}\tablespace}

\hrule\halign{&\vrule#&\strut\hskip0.2cm\hfil#\hfill\hskip0.2cm\cr
\tablespace
& sector && spin && states &\cr
\tablerule
& $ii$ && $(2,2)$ &&  U(4)$\times$U(4)$\times$U(4)$\times$U(4) & \cr
\tablespace
&    && $(2,1)$ &&  $2\times\left[ ({\rm adj},1,1,1)
+(1,{\rm adj},1,1)+
(1,1,{\rm adj},1)+ (1,1,1,{\rm adj}) \right]$ &\cr
\tablerule
& $i(i+2)$ && $(1,1)$ && $2\times\left[ (4,1,\o{4},1)
+(\o{4},1,{4},1)+
(1,4,1,\o{4})+(1,\o{4},1,{4}) \right]$ &\cr
\tablerule
& $i\tilde i$ && $(1,2)$ && $\left[(10+\o{10},1,1,1)
+(1,10+\o{10},1,1)+ 
(1,1,10+\o{10},1)+(1,1,1,10+\o{10}) \right]$ &\cr
\tablespace
& $$ && $(1,1)$ && $2\times\left[ (10+\o{10},1,1,1)+(1,10+\o{10},1,1)+
 (1,1,6+\o{6},1)+(1,1,1,6+\o{6}) \right]$ &\cr
\tablerule
& $i\widetilde{(i+2)}$ && $(2,1)$ && $\left[
 (4,1,4,1)+(\o{4},1,\o{4},1)+ 
 (1,4,1,4)+(1,\o{4},1,\o{4}) \right]$ &\cr
\tablerule
& $i\widetilde{(i+3)}$ && $(1,1)$ && $2\times\left[ (4,1,1,4)+
(\o{4},1,1,\o{4})+ (1,4,4,1)+(1,\o{4},\o{4},1) \right]$ &\cr
\tablespace}\hrule}}}}
\centerline{
\hbox{{\bf Table 4:}{\it ~~ Massless open string spectrum for model
II (orthogonal $D7$-branes).}}} 
\vskip 0.5cm 
\noindent
Again, the massless fermion spectrum cancels the non-factorizable
gravitational anomaly of the closed string spectrum, and ensure the
vanishing of irreducible gauge anomalies as well. 

The two D7-brane models are actually closely related (see also
\raldura\ for a similar construction). Suppose we add to the second
theory four bulk D7-brane anti-brane pairs that are parallel to the
$x_6-x_8$ plane, together with their images under $\Omega {\cal R}
I_4^L$, \ie\ together with another four D7-brane anti-brane pairs in
the bulk that are parallel to the $x_7-x_9$ plane. We can then
consider moving the anti-branes to the fixed planes at $x_6=0,x_8=1/2$
and $x_7=0,x_9=1/2$, respectively, while moving the branes to the
fixed planes at $x_6=x_8=1/2$ and $x_7=x_9=1/2$, respectively. A bulk 
brane carries twice the untwisted R-R charge of a fractional brane,
and therefore precisely changes the sign of the untwisted R-R charge
(if it is of opposite sign). Thus the above operation transforms model
II into model I.

\newsec{Non-BPS D8-branes}
\pano
Both of the above models contain massless scalars in the $i(i+2)$
sector. This sector consists of open strings that stretch between a
D7-brane and a parallel anti-D7-brane, for example between $D7_1$ and 
$\o{D7}_3$. The `tachyonic' gound state from the NS sector is
therefore invariant under the GSO-projection; its zero momentum and
winding component is removed by the orbifold projection, but states
with non-trivial winding and momentum survive. For the actual
configuration that we are considering (where the brane and the
anti-brane are separated by a finite distance along  the $x_6$
direction and where they carry a relative Wilson line in the $x_7$
direction), the lowest lying physical state is in fact massless. This
suggests that the corresponding scalars describe a marginal
transformation along which the D7-anti-D7-brane system can be deformed
into a non-BPS D8-brane that fills the space between the two D7-branes
\refs{\rsena,\rbgtwo,\rsenb,\rgabste} (see also
\refs{\rsenrev,\rlerdarev,\rgabrev} for a review of these matters).   

Actually, the relevant non-BPS D8-brane is not a conventional non-BPS
D-brane since the charge distribution at the end-points can not be
described in terms of constant Wilson lines. (This is basically a
consequence of the fact that the two D7-branes, $D7_1$ and $\o{D7}_3$,
out of which the D8-brane forms do not have the same Wilson line in
the $x_7$ direction.) In fact, the non-BPS D8-brane into which the
system decays carries a non-trivial {\it magnetic flux}. One way to
see this is to observe that the charge distribution requires that the
Wilson line in the $x_7$ direction depends nontrivially on $x_6$, \ie\ 
\eqn\ffield{ A_6=\half\,, \qquad A_7= x_6\,, \quad A_9=0\,,}
and therefore that the magnetic flux 
$\F_{67}=\partial_6 A_7 - \partial_7 A_6 = 1$. Alternatively, the 
correct charge distribution can be described by the superposition of a
conventional non-BPS D8-brane (where all eight twisted R-R charges are
$+$) together with a non-BPS D6-brane that stretches along $x_9$ (and
is localised at $x_6=\half$, $x_7=x_8=0$), both of whose twisted R-R 
charges are $-$. Since the magnitude of the twisted R-R charge at the
end of a non-BPS D6-brane is twice that of a non-BPS D8-brane, the
total charge of the configuration agrees then with that of the
D7-brane anti-brane configuration. On the other hand, the open string
between the D6-brane and the D8-brane contains a tachyon, and the
system decays (presumably) into a non-BPS D8-brane with magnetic flux.

It is not difficult to describe the non-BPS D8-brane with magnetic
flux in terms of boundary states. Let us consider the configuration
that is relevant to the previous discussion: it is localised at
$x_8=0$, and has magnetic flux $\F_{67}=1$. The boundary conditions
for the internal directions are then
\eqn\boundc{\eqalign{  
\partial_\sigma X_6 + \partial_\tau X_7 &=0\,, \cr
\partial_\sigma X_7 - \partial_\tau X_6 &=0\,, \cr
\partial_\sigma X_8 &=0\,,  \cr
\partial_\tau X_9 &=0 \,,\cr}} 
and the exponential of the bosonic oscillators is of the form
\eqn\oscib{  |B\rangle= {\rm exp}\left( \sum_{n} -{1\over n} \left(
               \alpha^6_{-n}\tilde \alpha^7_{-n} -
               \alpha^7_{-n}\tilde \alpha^6_{-n}-
                \alpha^8_{-n}\tilde \alpha^8_{-n}+
                 \alpha^9_{-n}\tilde \alpha^9_{-n} \right) \right) 
             |k,\omega\rangle }
and similarly for the fermions. Under $\Omega{\cal R} I_4^L$ this is
mapped to 
\eqn\oscic{  |\tilde B\rangle= {\rm exp}\left( \sum_{n} {1\over n} 
 \left(       \alpha^6_{-n}\tilde \alpha^7_{-n} -
               \alpha^7_{-n}\tilde \alpha^6_{-n}-
                \alpha^8_{-n}\tilde \alpha^8_{-n}+
                 \alpha^9_{-n}\tilde \alpha^9_{-n} \right) \right) 
             |k,\omega\rangle\,. }
The last boundary state describes a non-BPS D8-brane that is localised
at $x_9=0$, and that has magnetic flux $\F_{67}=-1$. This is indeed
the appropriate non-BPS D8-brane into which the combination of
$\widetilde{D7}_1$ and $\widetilde{\o{D7}}_3$ can decay.

Schematically speaking, the boundary state of the whole non-BPS
$D8$-brane has the form
\eqn\deight{ |D8\rangle=
          \left(|U,NS\rangle +|T,R\rangle \right) \,,}
where again the normalization is fixed by world-sheet duality to the
loop channel
\eqn\ann{ \cA= \C \int_0^\infty {dt\over t^4} {\rm Tr}
\left( {1+(-1)^F\, I_4 \over 4}\,
                e^{-2\pi t L_0  } \right)\,.}
In order to cancel the twisted sector tadpoles we need four such 
non-BPS $D8$-branes with parameters as shown in table~5, where again,
$\widetilde{D8}_i$ is the image of $D8_i$ under $\Omega{\cal R}I_4^L$.
\vskip 0.2cm
\centerline{\vbox{
\hbox{\vbox{\offinterlineskip
\def\tablespace{height2pt&\omit&&\omit&&\omit&&\omit&&
 \omit&\cr}
\def\tablerule{\tablespace\noalign{\hrule}\tablespace}

\hrule\halign{&\vrule#&\strut\hskip0.2cm\hfil#\hfill\hskip0.2cm\cr
\tablespace
& brane && location  && Wilson line && twisted sector && 
                $\F_{67}$ &\cr
\tablerule
& $D8_1$ && $x_8=0$  && $\theta_9=0$ &&
$(T_1+T_3)(T_1+T_3)+(T_4-T_2)(T_1+T_3)$ && $1$ &\cr
\tablespace
& $D8_2$ && $x_8={1\over 2}$  && $\theta_9={1\over 2}$ &&
$(T_1+T_3)(T_4-T_2)+(T_4-T_2)(T_4-T_2)$ && $1$ &\cr
\tablerule
& $\widetilde{D8}_1$ && $x_9=0$  && $\theta_8=0$ &&
$(T_1+T_2)(T_1+T_2)+(T_4-T_3)(T_1+T_2)$ && $-1$ &\cr
\tablespace
& $\widetilde{D8}_2$ && $x_9={1\over 2}$  && $\theta_8={1\over 2}$ &&
$(T_1+T_2)(T_4-T_3)+(T_4-T_3)(T_4-T_3)$ && $-1$ &\cr
\tablespace}\hrule}}}}
\centerline{
\hbox{{\bf Table 5:}{\it ~~ Non-BPS D8-branes.}}}
\vskip 0.5cm
\noindent
As was explained in some detail in \rbgkl, when dealing with branes
with background gauge fields two issues need special attention. Firstly,
the zero-mode spectrum of the open strings between two parallel
D-branes with background gauge flux changes to
\eqn\zerospec{  M^2={|r + U\, s|^2\over U_2}{T_2\over |n+ T\, m|^2}\,,}
where in our case $U=T=i$, $n=m=1$ and $n=-m=1$, respectively. Thus
compared to a D8-brane without magnetic flux one gets an extra factor
of one-half for the zero mode spectrum 
\eqn\zerospecb{  M^2={r^2 +s^2\over 2} \,,} 
leading to an extra factor of $\sqrt{2}$ in the normalization of the
D8-brane boundary states.\footnote{$^\ddagger$}{Using the Poisson
resummation formula, the factor of one-half in the zero mode spectrum
leads to an extra factor of two for the normalisation of the
tree-channel zero mode contribution; this requires an additional
factor of $\sqrt{2}$ for the boundary state.} In the overlap of two 
non-BPS D8-branes with $\F=1$ and $\F=-1$ (for which the winding and
momentum sum is absent), one therefore obtains an extra
multiplicity of two, implying that in loop channel every state in this
open string sector is two-fold degenerate. Both effects can easily be
seen in the T-dual picture involving branes at angles. A D-brane with
non-trivial gauge flux is generally mapped to a brane wrapping around
rational cycles of the $T^2$ different from the two fundamental ones;
apparently, this changes the zero mode spectrum. The second effect is
due multiple intersection points of D-branes intersecting at angles 
\refs{\rbgka,\rba,\rbgkb} (see also \rprad). In our case the T-dual
D-branes stretch along the main and the off-diagonal of the T-dual
torus. Therefore it is evident that we really get an extra factor of
two for open strings  between a non-BPS brane with $\F=1$ and a
non-BPS brane with $\F=-1$. 

Taking these two effects into account, the annulus amplitude becomes 
in tree channel
\eqn\anngbps{\eqalign{\widetilde{\cA}=\C\int_0^\infty dl\ 
{\rm N}^2 \Biggl\{
&\left( {f_3^8-f_4^8 \over f_1^8 } \right)_{\rm NSNS,U} 
\left(\sum_{m\in\ZZ} e^{-2 \pi l m^2} \right)^2 \Biggl[
             \left(\sum_{n\in\ZZ} e^{-\pi l n^2} \right)^2 \cr
 &\phantom{ffffffffffffffff} +
         \left(\sum_{n\in\ZZ} (-1)^n \,
          e^{-\pi l n^2}\right)^2
         \Biggr]   \cr
-8&\left( { f_2^4 f_3^4 - f_0^4 f_4^4 \over 
              f_1^4 f_4^4}\right)_{\rm RR,T} +
8\left({f_3^4 f_4^4 - f_4^4 f_3^4 \over f_1^4 f_2^4}\right)
\Biggr\}\,.\cr }}
The first two terms arise from the overlap of the boundary states for
two branes with identical magnetic fields, and the last term is the
untwisted part of the overlap of the  boundary states for two branes
with opposite magnetic fields. Since the twisted sector ground states
for two such branes are orthogonal to each other, the twisted sector
contribution vanishes. Upon a modular transformation this becomes in
loop channel  
\eqn\annfbps{\eqalign{\cA=\C\int_0^\infty {dt\over t^4}\ {\rm N}^2\ 
\Biggl\{ &\left( {f_3^8-f_2^8 \over f_1^8 }\right) 
\left(\sum_{m\in\ZZ} e^{-\pi t m^2} \right)^2 \Biggl[
\left(\sum_{n\in\ZZ} e^{-2\pi t n^2} \right)^2 \cr
& \phantom{fffffffffffffffff}  
+ \left(\sum_{n\in\ZZ} e^{-2\pi t \left(n+{1\over 2}\right)^2}\right)^2
  \Biggr]  \cr
-4&\left( { f_4^4 f_3^4 - f_0^4 f_2^4 \over f_1^4 f_2^4} \right) 
+4\left( {f_3^4 f_2^4 - f_2^4 f_3^4 \over f_1^4 f_4^4}\right) 
\Biggr\}\,, \cr}}
where we have used the Poisson resummation formula,
\eqn\poisson{\sum_{m\in\ZZ} e^{-\pi l (m/R)^2} = {R \over \sqrt{l}}
\sum_{n\in\ZZ} e^{-\pi (nR)^2 / l} \,.}
Note, that in \annfbps\ the contribution for two branes with
opposite magnetic fields and $(-1)^F I_4$ insertion vanishes
identically. This implies that $(-1)^F I_4$ acts with opposite
signs on the two-fold degenerate ground states in this sector.
Using the crosscap states defined in the appendix we can determine the
tree-channel M\"obius amplitude
\eqn\moibibps{ \widetilde{\cM}=\C\int_0^\infty dl\, 32\, {\rm N}\, 
   \left( {f_2^4 f_3^2 f_3 f_4-f_0^4 f_4^2 f_4 f_3 \over 
                  f_1^4 f_4^2 f_3 f_4}+
          {f_2^4 f_4^2 f_4 f_3 -f_0^4 f_3^2 f_3 f_4\over 
                  f_1^4 f_3^2 f_4 f_3}
\right) }
with argument $q=i{\rm exp}(-2\pi l)$. The tadpole cancellation
condition is therefore as before in \tada, \ie\ we need ${\rm N}=4$
non-BPS D8-branes of each kind. In loop channel, the M\"obius
amplitude is then 
\eqn\moibibpsl{ \cM=\C\int_0^\infty {dt\over t^4} \, 2\, {\rm N}\, 
  \left( e^{i{\pi\over 2}}\, 
{f_2^4 f_4^2 f_4 f_3-f_0^4 f_3^2 f_3 f_4 
             \over f_1^4 f_3^2 f_4 f_3} + 
   e^{-i{\pi\over 2}}\,  
    {f_2^4 f_3^2 f_3 f_4 -f_0^4 f_4^2 f_4 f_3
              \over f_1^4 f_4^2 f_3 f_4}
      \right)\,.}
Taking into account that $(-1)^F I_4$ acts with opposite signs on the
two-fold degenerate ground states of the open string between a brane
with $\F=1$ and one with $\F=-1$, and that at the massless level the
loop channel M\"obius amplitude \moibibpsl\ vanishes we derive the
massless open string spectrum presented in table~6. 
\vskip 0.8cm
\vbox{
\centerline{\vbox{
\hbox{\vbox{\offinterlineskip
\def\tablespace{height2pt&\omit&&\omit&&
 \omit&\cr}
\def\tablerule{\tablespace\noalign{\hrule}\tablespace}

\hrule\halign{&\vrule#&\strut\hskip0.2cm\hfil#\hfill\hskip0.2cm\cr
\tablespace
& sector && spin && states &\cr
\tablerule
& $ii$ && $(2,2)$ &&  U(4)$\times$U(4) & \cr
\tablespace
&    && $(2,1)$ &&  $4\times\left[ ({\rm adj},1)+(1,{\rm adj})\right]$
&\cr 
\tablerule
& $i(i+1)$ && $(1,1)$ && $2\times\left[ (4,\o{4})+(\o{4},{4}) \right]$
&\cr 
\tablerule
& $i\tilde i$ && $(1,2)$ && $\left[ (10+\o{10},1)+ (1,10+\o{10})
\right]$ &\cr 
\tablespace
&  && $(2,1)$ && $\left[ (6+\o{6},1)+ (1,6+\o{6})\right]$ &\cr
\tablespace
& $$ && $(1,1)$ && $2\times\left[ (10+\o{10},1)+(1,10+\o{10})+
(6+\o{6},1)+(1,6+\o{6}) \right]$ &\cr
\tablerule
& $i\widetilde{(i+1)}$ && $(2,1)$ && $\left[ (4,4)+
 (\o{4},\o{4}) \right]$ &\cr
\tablespace
&   && $(1,2)$ && $\left[ (4,4)+ (\o{4},\o{4}) \right]$ &\cr
\tablespace
& $$ && $(1,1)$ && $4\times\left[ (4,4)+
(\o{4},\o{4}) \right]$ &\cr
\tablespace}\hrule}}}}
\centerline{
\hbox{{\bf Table 6:}{\it ~~ Massless open string spectrum on non-BPS
$D8$-branes.}}}} 
\vskip 0.5cm 
\noindent
The spectrum of massless fermions cancels again the anomaly; in order
for this to work, it is important that the extra multiplicities
arising from double intersection points in the $D8$-$\widetilde{D8}$
sector is taken into account. This provides an independent
confirmation of our claim that the non-BPS D8-branes carry magnetic
flux.

\newsec{The configuration with $D9$-$\o{D9}$ branes}
\pano
The spectrum in table~6 contains massless scalars in the $(12)$
sector. They arise from open strings that stretch between the non-BPS
D8-branes localised at $x_8=0$, and those that are localised at
$x_8=\half$. These massless scalars describe the marginal deformation
along which the non-BPS D8-branes can be deformed into a $D9$-$\o{D9}$
brane pair.  

The $D9$-$\o{D9}$ branes into which the system can decay have to carry
magnetic flux in both the $x_6$-$x_7$ and the $x_8$-$x_9$
directions in order to reproduce again the correct twisted R-R sector 
charges. The corresponding boundary states are then characterised by
the equations
\eqn\boundc{\eqalign{  
\partial_\sigma X_6 + \F_{67}\partial_\tau X_7 &=0\,, \cr
\partial_\sigma X_7 - \F_{67}\partial_\tau X_6 &=0\,, \cr
\partial_\sigma X_8 + \F_{89}\partial_\tau X_9 &=0\,, \cr
\partial_\sigma X_9 - \F_{89}\partial_\tau X_8 &=0\,. \cr}}
For $\F_{67}=\F_{89}=+1$, the exponential of the bosonic oscillators is
then of the form
\eqn\oscib{  |B\rangle= {\rm exp}\left( \sum_{n} -{1\over n} \left(
               \alpha^6_{-n}\tilde \alpha^7_{-n} -
               \alpha^7_{-n}\tilde \alpha^6_{-n}+
               \alpha^8_{-n}\tilde \alpha^9_{-n} -
               \alpha^9_{-n}\tilde \alpha^8_{-n}  \right) \right) 
             |k,\omega\rangle }
and similarly for the fermions. Under the action of 
$\Omega{\cal R} I_4^L$ this boundary state is mapped to 
\eqn\oscic{  |\tilde B\rangle= {\rm exp}\left( \sum_{n} {1\over n} 
\left(         \alpha^6_{-n}\tilde \alpha^7_{-n} -
               \alpha^7_{-n}\tilde \alpha^6_{-n}+
              \alpha^8_{-n}\tilde \alpha^9_{-n} -
               \alpha^9_{-n}\tilde \alpha^8_{-n} \right) \right) 
             |k,\omega\rangle\,. }
This corresponds then to a $D9$-brane state with magnetic flux 
$\F_{67}=\F_{89}=-1$

Again schematically, the boundary states of the $D9$-brane and the
$\o{D9}$ brane have the form 
\eqn\nine{\eqalign{ &|D9\rangle=\left(
                      |U,NS\rangle+|U,R\rangle\right)
                +\left(|T,NS\rangle +|T,R\rangle \right) \cr
            &|\o{D9}\rangle=\left(
                      |U,NS\rangle-|U,R\rangle\right)
              -\left(|T,NS\rangle -|T,R\rangle \right)\,, \cr}}
where the normalisations are determined by world-sheet duality. The
open string loop amplitude of the open string stretched between the 
$D9$ and the $\o{D9}$ brane is then
\eqn\annnine{ \cA= \C\,\int_0^\infty {dt\over t^4} {\rm Tr}\,
\left( {1\over 2}\,{1-(-1)^F\over 2}\, {1-I_4 \over 2}\,
                e^{-2\pi t L_0  } \right) \,.}
In particular, since the $I_4$ projection appears now with the
opposite sign, the ground state tachyon is removed. Under 
$\Omega {\cal R} I_4^L$ the two boundary states \nine\ are now mapped 
to  
\eqn\tilnine{\eqalign{ &|\widetilde{D9}\rangle=
\left(|U,NS\rangle-|U,R\rangle\right)
-\left(|T,NS\rangle -|T,R\rangle \right) \cr
&|\widetilde{\o{D9}}\rangle=
\left(|U,NS\rangle+|U,R\rangle\right)
+\left(|T,NS\rangle +|T,R\rangle \right)\,. \cr}}
In particular, $\widetilde{D9}$ is an anti-brane and
$\widetilde{\o{D9}}$ a brane (since $\Omega\R$ maps a D9-brane into an
anti-brane and vice versa). Thus we are led to consider a
configuration of D9-branes and anti-branes as described in table~7.
\vskip 0.4cm
\centerline{\vbox{
\hbox{\vbox{\offinterlineskip
\def\tablespace{height2pt&\omit&&\omit&&\omit&&
 \omit&\cr}
\def\tablerule{\tablespace\noalign{\hrule}\tablespace}

\hrule\halign{&\vrule#&\strut\hskip0.2cm\hfil#\hfill\hskip0.2cm\cr
\tablespace
& brane && twisted sector && $\F_{67}$ &&  $\F_{89}$ &\cr
\tablerule
& $D9$ && 
$(T_1+T_3)(T_1+T_3)+(T_4-T_2)(T_1+T_3)
$ 
&& $1$ && $1$ &\cr
\tablespace
& && 
$+(T_1+T_3)(T_4-T_2)+(T_4-T_2)(T_4-T_2)$ 
&& && &\cr
\tablespace
& $\o{D9}$ &&
$(T_1+T_3)(T_1+T_3)+(T_4-T_2)(T_1+T_3)
$
&& $1$ && $1$ &\cr
\tablespace
& &&
$+(T_1+T_3)(T_4-T_2)+(T_4-T_2)(T_4-T_2)$ 
&& && &\cr
\tablerule
& $\widetilde{D9}$ && 
$(T_1+T_2)(T_1+T_2)+(T_4-T_3)(T_1+T_2)
$
&& $-1$ && $-1$ &\cr
\tablespace
& && 
$+(T_1+T_2)(T_4-T_3)+(T_4-T_3)(T_4-T_3)$ 
&& && &\cr
\tablespace
& $\widetilde{\o{D9}}$ &&
$(T_1+T_2)(T_1+T_2)+(T_4-T_3)(T_1+T_2)
$
&& $-1$ && $-1$ &\cr
\tablespace
& &&
$+(T_1+T_2)(T_4-T_3)+(T_4-T_3)(T_4-T_3)$ 
&& && &\cr
\tablespace}\hrule}}}}
\centerline{
\hbox{{\bf Table 7:}{\it ~~ $D9$-$\o{D9}$ branes.}}}
\vskip 0.5cm
\noindent
The annulus amplitude in tree channel then becomes
\eqn\anngbps{\eqalign{  \widetilde{\cA}=\C \int_0^\infty dl\ 
 {\rm N}^2 \Biggl\{
&2\, \left( {f_3^8-f_4^8\over f_1^8} \right)_{\rm NSNS,U} 
\left(\sum_{r,s\in\ZZ} e^{-2 \pi l {r^2+s^2-2\kappa rs\over 
       \sqrt{1-\kappa^2}}} \right)^2 \cr
-8&\left( { f_2^4 f_3^4 -f_0^4 f_4^4 
\over f_1^4 f_4^4}\right)_{\rm RR,T} +
8\left( { f_3^4 f_4^4 - f_4^4 f_3^4 -f_2^4 f_0^4 +f_0^4 f_2^4 \over 
f_1^4 f_2^4}\right)
\Biggr\}\,,\cr }}
where we have introduced, for later convenience, 
$\kappa=\kappa_{67}=\kappa_{89}$ to denote the tilt of the $x_6-x_7$
and the $x_8-x_9$ torus. Using the crosscap states defined in the
appendix we can also determine the tree-channel M\"obius
amplitude\footnote{$^\dagger$}{For $\kappa\ne 0$, the crosscap states
are modified in the obvious way; this does not effect the Klein bottle
amplitude however.}
\eqn\moibibps{ \widetilde{{\cal M}}=\C\int_0^\infty dl\, 64\, {\rm N}\, 
\left( {f_2^4 f_3^2 f_4^2 - f_0^4 f_4^2 f_3^2 \over f_1^4 f_3^2 f_4^2}
        \right)\,, }
and thus read off the tadpole cancellation condition. This turns out
to be the same as \tada, independent of $\kappa$, and we thus have to
choose ${\rm N}=4$. The massless spectrum depends on the other hand on
$\kappa$ (as we shall discuss below); for $\kappa=0$ it is shown in
table~8. As before, the massless fermions cancel the non-factorizable
gravitational anomaly.
\vskip 0.8cm
\centerline{\vbox{
\hbox{\vbox{\offinterlineskip
\def\tablespace{height2pt&\omit&&\omit&&
 \omit&\cr}
\def\tablerule{\tablespace\noalign{\hrule}\tablespace}

\hrule\halign{&\vrule#&\strut\hskip0.2cm\hfil#\hfill\hskip0.2cm\cr
\tablespace
& sector && spin && states &\cr
\tablerule
& $ii$ && $(2,2)$ &&  U(4)$\times$U(4) & \cr
\tablespace
&   && $(2,1)$ &&  $2\times\left[ 
({\rm adj},1)+(1,{\rm adj})\right]$ &\cr
\tablerule
& $i(i+1)$ && $(1,1)$ && $8\times\left[ 
(4,\o{4})+(\o{4},{4}) \right]$ &\cr
\tablespace
& && $(2,1)$ && $2\times\left[ (4,\o{4})+(\o{4},{4}) \right]$ &\cr
\tablerule
& $i\tilde i$ && $(2,1)$ && $2\times \left[ 
(6+\o{6},1)+ (1,6+\o{6}) \right]$ 
&\cr
\tablespace
& $$ && $(1,1)$ && $2\times\left[ (10+\o{10},1)+(1,10+\o{10})+
(6+\o{6},1)+(1,6+\o{6}) \right]$ &\cr
\tablerule
& $i\widetilde{(i+1)}$ && $(1,2)$ && $2\times\left[ (4,4)+
 (\o{4},\o{4}) \right]$ &\cr
\tablespace
& $$ && $(1,1)$ && $4\times\left[ (4,4)+
(\o{4},\o{4}) \right]$ &\cr
\tablespace}\hrule}}}}
\centerline{
\hbox{{\bf Table 8:}{\it ~~ Massless open string spectrum on 
$D9$-$\o{D9}$ branes.}}}

\newsec{The configuration with diagonal $D7$-branes}
\pano
In the previous sections we have described a number of different
tadpole cancelling configurations that are related to each other by
standard deformations. There exists one other interesting brane
distribution that is not so obviously related to these configurations,
but that is relevant for the stability analysis of the theory. This
configuration consists of $D7$-branes and anti-branes that stretch
diagonally across the tori. If we denote by $y_1$ and $y_2$ the
coordinate along the main diagonal of the two $T^2$s, this brane
configuration is described in table~9 (see also figure~4). 
\vskip 0.2cm
\centerline{\vbox{
\hbox{\vbox{\offinterlineskip
\def\tablespace{height2pt&\omit&&\omit&&\omit&&
 \omit&\cr}
\def\tablerule{\tablespace\noalign{\hrule}\tablespace}

\hrule\halign{&\vrule#&\strut\hskip0.2cm\hfil#\hfill\hskip0.2cm\cr
\tablespace
& brane && location  && Wilson line && twisted sector  &\cr
\tablerule
& $D7$ && $y_1=y_2=0$  && $\theta_1=\theta_2=0$ &&
$(T_1+T_4)(T_1+T_4)$ &\cr
\tablespace
& $\o{D7}$ && $y_1=y_2={1\over 2}$  && $\theta_1=\theta_2={1\over 2}$ 
&& $(T_3-T_2)(T_3-T_2)$  &\cr
\tablespace}\hrule}}}}
\centerline{
\hbox{{\bf Table 9:}{\it ~~ Diagonal  $D7$-$\o{D7}$ branes.}}}
\vskip 0.5cm
\fig{\ The configuration with diagonal
$D7$-branes.}{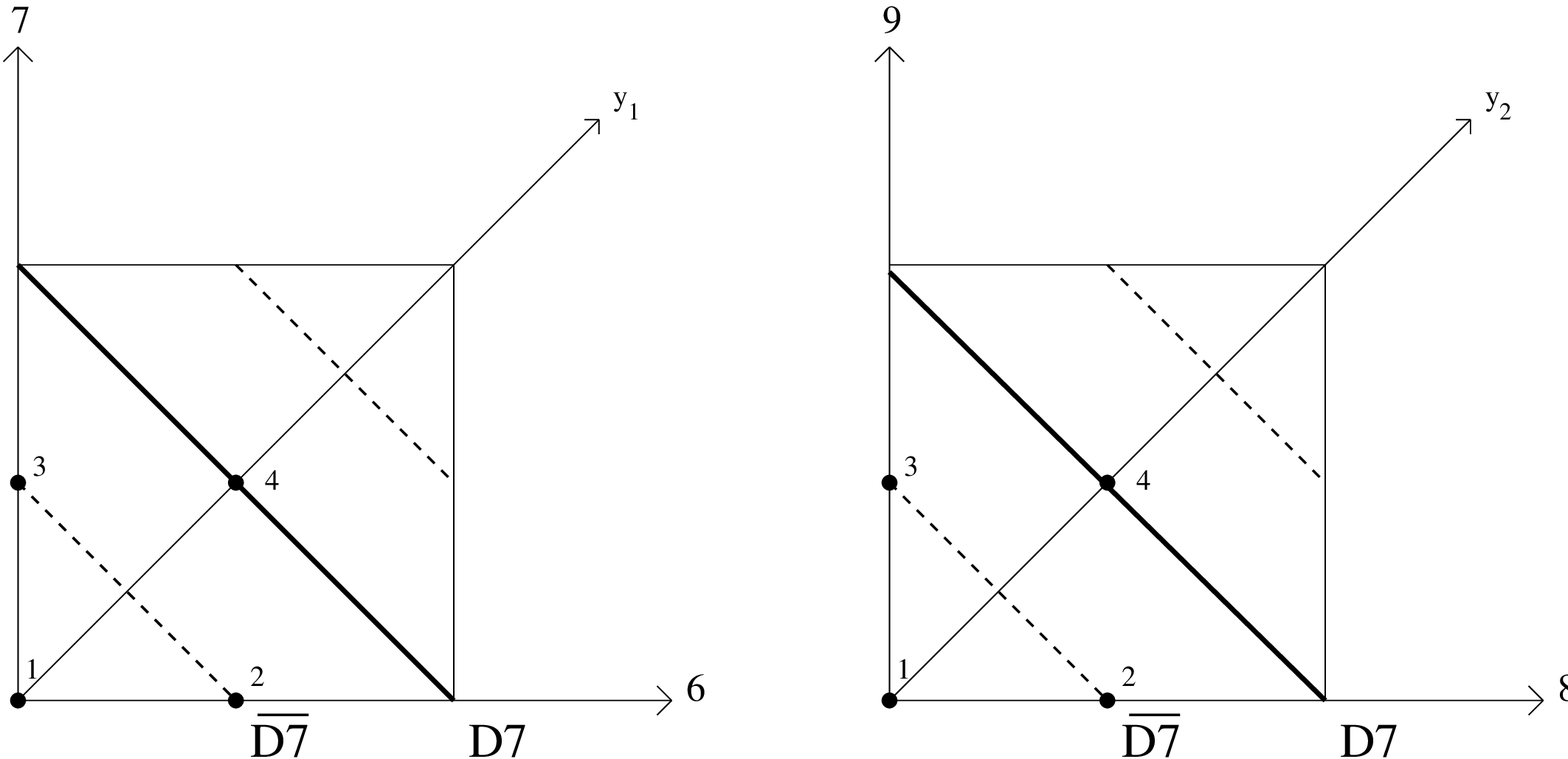}{14truecm} 
{From} the point of view of the boundary states, it is actually not
surprising that this configuration also cancels the tadpoles. The
total crosscap state $|C\rangle=|C_L\rangle + |C_R\rangle$ only has
twisted R-R charge at eight of the sixteen corners, since the
contributions from $|C_L\rangle$ and $|C_R\rangle$ cancel at the other
eight fixed points. (This can be directly seen from \twista.) Where
the contributions add, the charge is twice as large as before, and we
therefore expect that we shall need {\it eight} D7-brane-anti-brane
pairs of the above type. This is indeed correct: the tree channel
annulus amplitude of the above brane system is
\eqn\anngdia{\eqalign{  \widetilde{\cA}=\C\int_0^\infty &dl\ {\rm N}^2
\Biggl\{ {1\over 2} \left({1-\kappa\over
  1+\kappa}\right)\, \left( {f_3^8 -f_4^8 
   \over f_1^8 }\right)_{\rm NSNS,U} 
  \Biggl[ \left( \sum_{r\in\ZZ} e^{-2 \pi l \sqrt{1-\kappa\over
  1+\kappa}\, r^2} \right)^4  \cr
&\phantom{hhhhhhhhhhhhhhhhhhhhhhhhhhhhhhhhhhh}  
 +  \left( \sum_{r\in\ZZ} (-1)^r\, e^{-2 \pi l 
    \sqrt{1-\kappa\over 1+\kappa}\, r^2} \right)^4\Biggr] \cr
&- {1\over 2} \left({1-\kappa\over
  1+\kappa}\right)\,
\left( {f_2^8-f_0^8\over f_1^8  }\right)_{\rm RR,U} 
\Biggl[ \left( \sum_{r\in\ZZ} e^{-2 \pi l \sqrt{1-\kappa\over
  1+\kappa}\, r^2} \right)^4 \cr
&\phantom{hhhhhhhhhhhhhhhhhhhhhhhhhhhhhhhhhhh} 
-  \left( \sum_{r\in\ZZ} (-1)^r\, e^{-2 \pi l 
\sqrt{1-\kappa\over 1+\kappa}\, r^2} \right)^4\Biggr] \cr
&+ 2\left( {f_3^4 f_2^4 - f_4^4 f_0^4 - f_2^4 f_3^4 + f_0^4 f_4^4\over
      f_1^4 f_4^4} \right)_{\rm NSNS-RR,T} \Biggr\}
\,,\cr }}
where we have again performed the calculation for general tilt
parameter $\kappa$. On the other hand, the contribution to the twisted
R-R tadpole of the tree channel M\"obius amplitude is
\eqn\moebdiag{ \widetilde{\cM}=\C\int_0^\infty dl\ 32\, {\rm N}\ 
      \left(    { \th{{1/2}}{0}^2 \th{0}{1/4}^2
\over \eta^{6}\, \th{0}{{1/4}}^2 }\right)_{\rm RR,T} \,}
with
\eqn\defth{ {\th{\alpha}{\beta} \over \eta}(q)=e^{2\pi i \alpha\beta}\,
             q^{{\alpha^2\over 2} - {1\over 24}}\, 
             \prod_{n=1}^\infty \left(
     \left( 1+q^{n-{1\over 2}+\alpha}\, e^{2\pi i\beta}\right)
     \left( 1+q^{n-{1\over 2}-\alpha}\, e^{-2\pi i\beta}\right) 
  \right)}
and
\eqn\etadef{ \eta(q) = q^{1\over 24} \, \prod_{n=1}^{\infty}
      (1 - q^n) \,.}
Thus the tadpole cancellation condition becomes
\eqn\taddia{   2\, {\rm N}^2 - 32\, {\rm N} +128
       =2\, ({\rm N}-8)^2=0 \,.}
This requires indeed ${\rm N}=8$ D7-branes and $\o{D7}$-branes.
For $\kappa=0$, the massless open string spectrum that follows from
this configuration is given in table~10.
\vskip 0.8cm
\centerline{\vbox{
\hbox{\vbox{\offinterlineskip
\def\tablespace{height2pt&\omit&&\omit&&
 \omit&\cr}
\def\tablerule{\tablespace\noalign{\hrule}\tablespace}

\hrule\halign{&\vrule#&\strut\hskip0.2cm\hfil#\hfill\hskip0.2cm\cr
\tablespace
& sector && spin && states &\cr
\tablerule
& $ii$ && $(2,2)$ &&  Sp(8)$\times$SO(8) & \cr
\tablespace
&    && $(2,1)$ &&  $2\times\left[ (28,1)+(1,28)\right]$ &\cr
\tablerule
& $i\o{i}$ && $(1,1)$ && $8\times (8,8)$ &\cr 
\tablespace}\hrule}}}}
\centerline{
\hbox{{\bf Table 10:}{\it ~~ Massless open string spectrum on diagonal
$D7$-$\o{D7}$ branes.}}}
\vskip 0.5cm 
\noindent
In this case there are $112$ fermions of spin $(2,1)$, and this
cancels indeed the non-factorizable gravitational anomaly. There
also exists the configuration for which the $D7$-branes and
anti-branes stretch along the off-diagonal; the analysis of this case
is identical to the above.

\newsec{Regimes of stability}
\pano
In the previous sections we have discussed a number of different
tadpole cancelling configurations, all of which are free of tachyons
at the point in moduli space where the torus is an orthogonal
$SU(2)^4$ torus. In this section we want to explore which of
these configurations is stable at a more general point in moduli
space; we shall only consider a one-parameter subspace of the 
six-dimensional moduli space of tori (that was described in
section~2), but it is clear that at least the essential arguments
and observations will hold more generally. 

Let us consider the deformation of the torus where we tilt both the
(67) and the (89) torus, and let us, for simplicity, assume that
$\kappa=\kappa_{67}=\kappa_{89}$. As we increase $\kappa$, the
distance between a brane at $x_i=0$, and an anti-brane at $x_i=1/2$ is
reduced; in particular, if the `tachyonic' open string between these
two branes is massless for $\kappa=0$, it will become tachyonic for
$\kappa\neq 0$. This is precisely what happens for the (13) string of 
the original orthogonal $D7$-$\o{D7}$ brane system, and therefore the
marginal perturbation along which this system can be deformed into a
system of non-BPS D8-branes becomes relevant. The argument also
applies to the $i(i+1)$ sector of the non-BPS D8-brane system, and the
marginal deformation of this system into the $D9$-$\o{D9}$ brane
system becomes also relevant. Thus, for $\kappa\neq 0$, either of the
configurations described in sections~3 and 4 decays into the
$D9$-$\o{D9}$ brane configuration described in section~5
\eqn\dec{D7-\o{D7}\; ({\rm Section~3}) 
\longrightarrow 
     {\rm non-BPS-}D8\; ({\rm Section~4}) 
\longrightarrow
     D9-\o{D9}\; ({\rm Section~5})\,.}
On the other hand, from the point of view of the $D9$-$\o{D9}$ system,
the massless scalars in the $i(i+1)$ sector of table~8 (that describe
the marginal deformation back to the non-BPS $D8$-brane system) become
massive for $\kappa\neq 0$. Indeed, the mass formula for the KK states
on each $T^2$ is given by
\eqn\massfo{\eqalign{  
M^2&={T_2\over |n+T\, m|^2}\, {|r-U\, s|^2\over U_2 } - \half
    \cr
   &={r^2+s^2-2\kappa rs\over 2 \sqrt{1-\kappa^2}} -\half\,. \cr}}
For $\kappa=0$, the massless states in the $i(i+1)$ sector correspond
to states with $r=\pm 1, s=0$ or $r=0, s=\pm 1$. If $\kappa\neq 0$,
these states become indeed massive. (This also follows directly from
the loop amplitude 
\eqn\annfnine{\eqalign{  {\cA}=\C \int_0^\infty {dt\over t^4} 
 {\rm N}^2\, \Biggl\{
&\left( {f_3^8-f_2^8\over f_1^8} \right) 
\left(\sum_{r,s\in\ZZ} e^{- \pi t {r^2+s^2-2\kappa rs\over 
       \sqrt{1-\kappa^2}}} \right)^2 \cr
-4&\left( { f_4^4 f_3^4 -f_0^4 f_2^4 \over f_1^4 f_2^4}\right) +
4\left( { f_3^4 f_2^4 - f_4^4 f_0^4 - f_2^4 f_3^4 +f_0^4 f_4^4 \over 
f_1^4 f_4^4}\right)
\Biggr\}\,,\cr }}
that can be obtained by a modular transformation from the tree channel
amplitude \anngbps.)

For $0<\kappa<3/5$, the open string spectrum of the $D9$-$\o{D9}$
brane system is tachyon free, and this should imply that the
configuration is indeed stable. At $\kappa=3/5$, however, the states 
with  $(r,s)=\pm(1,1)$ become massless, and for $3/5<\kappa<1$, in
fact tachyonic. This implies that the stable configuration in this
domain is described by another system. Intuitively, these tachyons
arise because for $\kappa>3/5$, the D9-branes are  stretched too much
along the main diagonal direction of the torus, and it becomes
energetically preferable to decay into two non-BPS D8-branes that
stretch along the off-diagonal direction. The corresponding brane
configuration can be constructed, but it is also always unstable to
decay into the diagonal D7-brane system that we described in
section~6. Thus we find that the stable configuration for
$3/5<\kappa<1$ is described by diagonal $D7$-branes and anti-branes! 

Conversely, we can analyze the stability of the diagonal $D7$-brane
system for all values of $\kappa$. The mass formula for KK and winding
states for the off-diagonal branes on each $T_2$ is given by 
\eqn\masskappa{  M^2={r^2+s^2\over 2} \sqrt{1+\kappa\over 1-\kappa} -
\half\,.} 
For $\kappa\ge 0$ the system is therefore non-tachyonic and
the massless states in the $i\o{i}$ sector become massive for
$\kappa>0$. For $\kappa\le 0$ the open string between the $D7$ and the
$\o{D7}$  brane becomes tachyonic and the system decays into a $D7$
and a $\o{D7}$ along the off-diagonal. 

It is worth mentioning that the system does not develop a marginal
deformation (let alone a relevant deformation) for $\kappa=3/5$; this
suggests that the configuration of diagonal $D7$ branes and
anti-branes is actually stable for all values of $\kappa$, and that 
the $D9$-brane anti-brane system is only metastable (for
$|\kappa|<3/5$). 
\vskip 0.3cm

Finally, we have to address the question of what the effect of the
dynamically generated potential for $\kappa$ is. In the $D9$-brane
configuration, the NS-NS tadpoles do not depend on $\kappa$ (as
follows from \anngbps), and the first $\kappa$ dependent contribution
to the potential arises at one loop. Since there exist NS-NS tadpoles,
we are not really sitting in a string theory vacuum and the background
fields get modified by the Fischler-Susskind mechanism. Nevertheless,
since the $\kappa$ dependence of $\Lambda_{1-loop}$ only arises via
the Kaluza-Klein and winding modes, we are confident that qualitative
features of the $\kappa$ dependence can be reliably extracted from
the one-loop partition functions computed above. In particular
we compute 
\eqn\csom{  \Lambda_{1-loop}(\kappa)-\Lambda_{1-loop}(0)=
             -\cA(\kappa)+\cA(0) ,}
where in fact the contributions from the torus, the Klein-bottle
and the M\"obius strip vanish and only the first term in \annfnine\ 
contributes. Numerically evaluating \csom\ yields the curve depicted 
in figure 5.  
\fig{\ $\Lambda_{1-loop}(\kappa)-\Lambda_{1-loop}(0)$ for the
$D9$-brane system.}{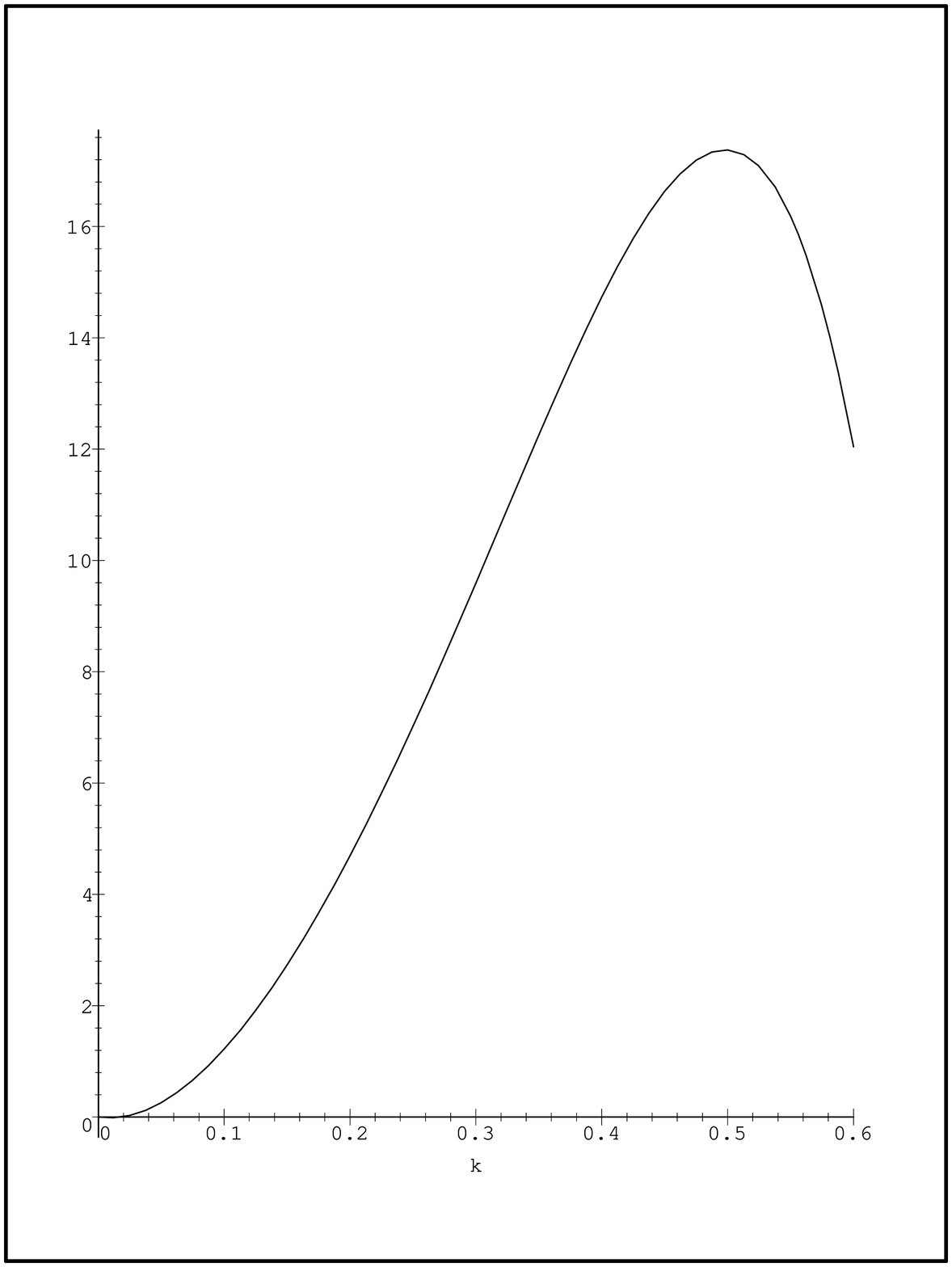}{12truecm}
It follows from this result that to one-loop, $\kappa=0$ is a 
stable minimum. It is separated by a finite potential barrier from the
configuration of diagonal $D7$-branes (into which it can decay at
$\kappa=3/5$). 

For the configuration of diagonal $D7$-branes, the NS-NS tadpole 
depends on $\kappa$, thus giving rise to a tree level contribution to 
the potential (see figure~6)
\eqn\pottree{   V(\Phi,\kappa)\sim e^{-\Phi}\, c\, 
\sqrt{1-\kappa\over  1+\kappa}\, {\rm N}\, +V_{1-loop}+\ldots \,.}
The potential is minimized in the singular limit $\kappa=1$; this
simply expresses the fact that the tension of the $D7$-branes pulls
the two sides of each $T^2$ together. However, the point $\kappa=1$ is
infinitely far away in moduli space and does not represent an actual
decay mode. Moreover, it might also happen that higher loop or
non-perturbative contributions to the potential stabilize $\kappa$
at a finite value $0\le \kappa<1$. The actual form of these
contributions is however beyond computational control.
\fig{\ The tree level potential for $\kappa$ for the configuration
with diagonal $D7$-branes.}{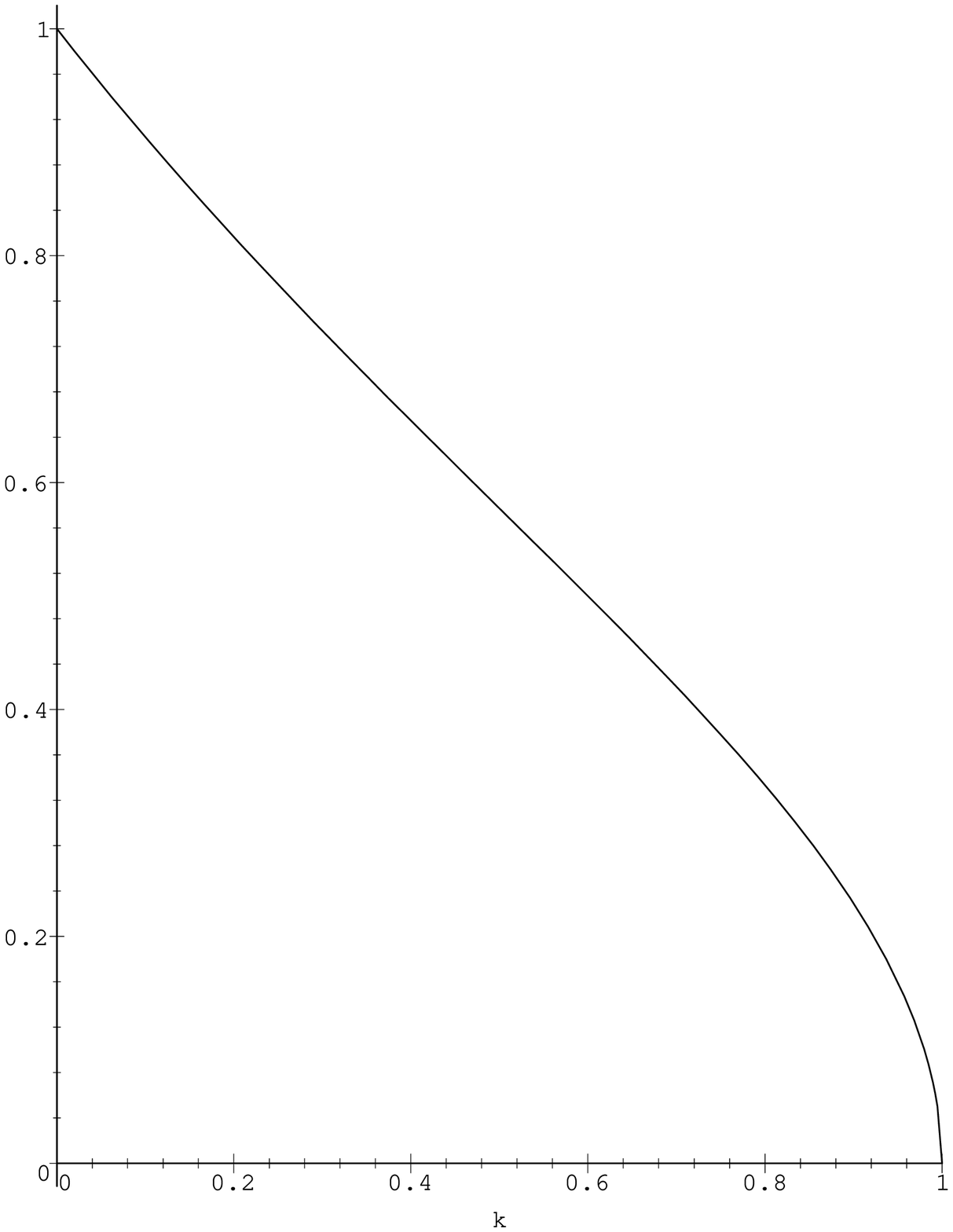}{8truecm}

\newsec{Conclusion}
\pano
In this paper we have constructed a new kind of asymmetric
orientifold in six dimensions which is supersymmetric in the bulk
and non-supersymmetric on the branes. Tadpole cancellation naively led
to the introduction of pairs of fractional $D7$ and $\o{D7}$ branes, 
which were localized on different fixed points, thus preventing the
development of tachyons.  However, by turning on some of the closed
string moduli the configuration of $D7$-$\o{D7}$-branes became
unstable and via non-BPS D8-branes eventually decayed into pairs of
$D9$-$\o{D9}$ branes with magnetic flux. For this configuration we
computed the one-loop cosmological constant and found that the system
is stabilized on the $SU(2)^4$ torus. However, this configuration is
only metastable, and it is separated by a finite energy barrier from
the stable system consisting of diagonal D7-branes and anti-branes.  

One of the lessons of this analysis is that models that are
tachyon-free at first sight (such as the configuration described in
section~3) can in fact be highly unstable, and may well decay into
more exotic brane configurations. In fact, in order to get control
over the stable configurations, it is important to analyze the various
closed string moduli in some detail. As far as we are aware, the
present paper is the first example where this has been done to any
degree. 

Most of the configurations that we found were at criticality in the
sense that some of the tachyonic modes were precisely
massless. Non-supersymmetric configurations at criticality sometimes
lead to precise Bose-Fermi degeneracy in the open string spectrum 
\rgabsen. However, for the class of configurations that we considered,
here, this did not occur; it would be interesting to find a model
where Bose-Fermi degeneracy is realized in the supersymmetry breaking
open string sector.

Finally, if we compactify the six-dimensional model on a further
$T^2$ and T-dualise this torus, we obtain a four-dimensional model,
for which the latter torus can be made large, thus leading to some
brane world scenario. It would therefore be interesting to generalize
our approach directly to four-dimensional models.

\vskip 1cm

\centerline{{\bf Acknowledgements}}\pano
C.A. is supported by the ``Marie-Curie'' fellowiship
HPMF-CT-1999-00256, in part by the EEC-IHP ``Superstring Theory''
contract HPRN-CT-2000-00122, in part by the EEC contract
ERBFMRX-CT96-0090, in part by the EEC contract ERBFMRX-CT96-0045 and
in part by the INTAS project 991590. 
R.B. is supported in part by the EEC contract
ERBFMRX-CT96-0045, and thanks the Erwin Schr\"odinger Institute for
Mathematical Physics, where part of this work has been
performed. M.R.G. is supported by a  Royal Society University Research
Fellowship, and acknowledges partial support from the PPARC SPG
programme, ``String Theory and Realistic Field Theory'',
PPA/G/S/1998/00613.   

We would like to thank Gary Shiu for collaboration during an early
stage of this work. We also thank Massimo Bianchi, Dieter L\"ust, Jos\'e
F. Morales and Augusto Sagnotti for helpful discussions.  

\vskip 1cm

\appendix{A}{The crosscap states}

Let us recall from \cross\ that the crosscap states 
$|\Omega{\cal R} I_4^{L,R}\rangle$ satisfy the equation
\eqn\crossone{ \left( X^\mu(\sigma,0) - {\cal R}\, I_4^{L,R}\, 
  X^\mu(\sigma+\pi,0)\right) |\Omega{\cal R}I_4^{L,R}\rangle=0\,,} 
where $\mu=2,\ldots,9$. In terms of the oscillator modes, this can be
rewritten as 
\eqn\osci{\eqalign{ \left( \alpha^m_r \pm \epsilon_m e^{-i\pi r}
  \tilde\alpha^m_{-r} \right) |\Omega{\cal R} I_4^{L,R}\rangle&=0, 
        \quad\quad{\rm for}\ m\in\{6,7,8,9\} \cr
       \left( \alpha^m_r \mp \epsilon_m e^{i\pi r}
  \tilde\alpha^m_{-r} \right) |\Omega{\cal R} I_4^{L,R}\rangle&=0,
        \quad\quad{\rm for}\ m\in\{6,7,8,9\} \cr
  \left( \alpha^\mu_n + (-1)^n
      \tilde\alpha^\mu_{-n} \right) |\Omega{\cal R} I_4^{L,R}\rangle&=0,
  \quad\quad{\rm for}\ \mu\in\{2,3,4,5\}\,, \cr}} 
where
\eqn\eps{   \epsilon_m=\cases{ +1 & for $m=6,8$ \cr
                               -1 & for $m=7,9$. \cr} }
The upper sign in \osci\ corresponds to 
$|\Omega{\cal R} I_4^{L}\rangle$, whereas the lower sign refers to 
$|\Omega{\cal R} I_4^{R}\rangle$. The first two conditions in \osci\
are compatible only for $r\in\ZZ+1/2$, confirming our general
observation that the crosscap state is a coherent state in the $I_4$
twisted sector. The crosscap conditions \osci\ also give rise to a
relation among the zero-modes which, as expected, can only be solved 
trivially. The conditions that arise for the fermionic modes are
similar 
\eqn\ferm{\eqalign{ \left( \psi^m_r\pm i\epsilon_m\,\eta\, e^{-i\pi r} 
  \tilde\psi^m_{-r} \right) |\Omega{\cal R} I_4^{L,R},\eta\rangle&=0,
      \quad\quad{\rm for}\ m\in\{6,7,8,9\},   \cr
    \left( \psi^\mu_r + i\, \eta\, e^{-i\pi r}
  \tilde\psi^\mu_{-r} \right) |\Omega{\cal R} I_4^{L,R},\eta\rangle&=0,
    \quad\quad{\rm for}\ \mu\in\{2,3,4,5\}\,, \cr } }
where, as usual, $\eta=\pm 1$ gives rise to the different spin
structures. The solution to these equations is given by
\eqn\crossb{\eqalign{|\Omega{\cal R} I_4^{L,R},\eta\rangle
   =\M \;{\rm exp}
 \biggl(&-\sum_{\mu=2}^5 \sum_{n\in\ZZ} {(-1)^n\over n} \alpha^\mu_{-n}
               \tilde\alpha^\mu_{-n} + 
       \sum_{m\in\{6,8\} } \sum_{r\in\ZZ+{1\over 2}} {e^{\pm i\pi r}
              \over r} \alpha^m_{-r}
               \tilde\alpha^m_{-r} \cr
            &+ \sum_{m\in\{7,9\} } \sum_{r\in\ZZ+{1\over 2}} 
               {e^{\mp i\pi r} \over r} \alpha^m_{-r}
               \tilde\alpha^m_{-r}  \cr
            &+ i \eta \biggl[-\sum_{\mu=2}^5 \sum_{r} e^{-i\pi r} 
               \psi^\mu_{-r} \tilde\psi^\mu_{-r} \mp
            \sum_{m\in\{6,8\} } \sum_{r} e^{-i\pi r}
               \psi^m_{-r}   \tilde\psi^m_{-r} \cr
            &\qquad \pm \sum_{m\in\{7,9\} } \sum_{r} 
               e^{- i\pi r} \psi^m_{-r} \tilde\psi^m_{-r}
               \biggr] \biggr)|T_{L,R},\eta\rangle\,, \cr}}
where the overall normalization $\M$ is determined by worldsheet
duality. The moding of the fermions $\psi^\mu_r$ and
$\tilde\psi^\mu_r$ depends on whether we are considering the twisted  
NS-NS or the twisted R-R sector of the theory. 

The ground states in the twisted sectors are constrained by the
conditions that arise from the fermionic zero modes in \ferm. In the
twisted R-R sector, the theory has only fermionic zero modes in the
directions unaffected by the orientifold, and therefore the standard
argument applies (see for example \rpolcai). On the other
hand, there are fermionic zero modes for $\psi^m_0$ with $m=6,7,8,9$
in the twisted NS-NS sector, and they give rise to the conditions
\eqn\zero{\eqalign{\psi^m_+ |T_{L},+\rangle 
& =0 \qquad \hbox{for $m=6,8$} \cr
\psi^m_- |T_{L},+\rangle 
& =0 \qquad \hbox{for $m=7,9$}}}
and similarly for $T_R$, 
\eqn\zeroone{\eqalign{\psi^m_- |T_{R},+\rangle 
& =0 \qquad \hbox{for $m=6,8$} \cr
 \psi^m_+ |T_{R},+\rangle 
& =0 \qquad \hbox{for $m=7,9$.}}}
Here we have defined 
\eqn\plusm{\psi^m_{\pm} = {1 \over \sqrt{2}} 
       \left(\psi^m_0 \pm i \tilde\psi^m_0 \right)\,.} 
For the IIB orbifold under consideration, the GSO-projection in the
twisted NS-NS sector is given by  
\eqn\mata{ {1\over 4} \left(1+(-1)^F\right) 
                \left(1+(-1)^{\widetilde{F}}\right) \,,}
where the two operators $(-1)^F$ and $(-1)^{\widetilde{F}}$ are defined
by
\eqn\matb{\eqalign{(-1)^F & = \prod_{m=6}^{9} \sqrt{2} \psi^m_0 
= \prod_{m=6}^{9} (\psi^m_+ + \psi^m_- ) \cr
(-1)^{\widetilde{F}} & = \prod_{m=6}^{9} \sqrt{2} \tilde\psi^m_0 
= \prod_{m=6}^{9} (\psi^m_+ - \psi^m_- )\,.}}
Thus if we define
\eqn\matc{|T_L,-\rangle 
    = \psi^6_- \psi^7_+ \psi^8_- \psi^9_+ |T_L,+\rangle \,,} 
from which it follows that 
\eqn\matd{|T_L,+\rangle 
    = \psi^6_+ \psi^7_- \psi^8_+ \psi^9_- |T_L,-\rangle \,,} 
we have that 
\eqn\mate{\eqalign{(-1)^F |T_L,\pm\rangle & = |T_L,\mp\rangle \,, \cr
(-1)^{\widetilde{F}} |T_L,\pm\rangle & = |T_L,\mp\rangle\,,}}
and therefore
\eqn\ground{ \left(|T_L,+\rangle + |T_L,-\rangle \right)}
is a GSO-invariant state. Since the GSO-operators act in the standard
way on the oscillator exponential, this implies that 
\eqn\total{ |\Omega\R I_4^L \rangle \equiv 
       \left(|\Omega \R I_4^L,+\rangle 
               +  |\Omega \R I_4^L,-\rangle \right)}
is GSO-invariant. The analysis for $\Omega \R I_4^R$ is identical
since the comparison of \zero\ with \zeroone\ implies that we can
define $|T_R,+\rangle=|T_L,-\rangle$ and
$|T_R,-\rangle=|T_L,+\rangle$. This analysis applies separately for
each twisted sector of the theory.

The actual crosscap states also have to be invariant under the
orientifold projection $\Omega \R I_4^L$ (which generates the whole
orientifold group). Since these crosscap states are effectively
O7-planes, the invariance under $\Omega \R$ is familiar. In order to
understand this more explicitly (compare \rbgone\ for a similar
analysis), we recall that $\Omega$ acts on the fermionic modes as
\eqn\matf{\eqalign{ \Omega \psi^m_r \Omega^{-1} & = \tilde \psi^m_r \cr
\Omega \tilde\psi^m_r \Omega^{-1} & = -\psi^m_r \,,}}
so that 
\eqn\matg{\eqalign{ \Omega \psi^m_+ \Omega^{-1} & = -i \psi^m_+ \cr
\Omega \tilde\psi^m_- \Omega^{-1} & = +i \psi^m_- \,.}}
If we denote by $|C9,\eta\rangle$ the ground state that satisfies 
\ferm\ without $\epsilon_m$, then $\Omega$ is defined to satisfy
$\Omega |C9,\eta\rangle =|C9,\eta\rangle$. Since
$|T_L,+\rangle=\psi^7_- \psi^9_- |C9,+\rangle$, it then follows that
$|T_L,\eta\rangle$ has eigenvalue $-1$ under the action of $\Omega$,
\ie\ $\Omega |T_L,\eta\rangle = - |T_L,\eta\rangle$. On the other
hand, $\R$ acts on the ground states as 
\eqn\math{\R= 4 \psi^7_0 \tilde\psi^7_0  \psi^9_0 \tilde\psi^9_0
       = - (\psi^7_+ + \psi^7_-) (\psi^7_+ - \psi^7_-) 
           (\psi^9_+ + \psi^9_-) (\psi^9_+ - \psi^9_-) \,,}
and thus $\R |T_L,\eta\rangle = - |T_L,\eta\rangle$. This implies that
$\Omega\R$ leaves $|T_L,\eta\rangle$ invariant. Again, the action on
the oscillator states is trivial, and therefore also 
$|\Omega \R I_4^L\rangle$ is invariant under $\Omega\R$. The same 
argument obviously also applies to $|\Omega \R I_4^R\rangle$.

\listrefs

\bye